# Electron-phonon coupling and mobility modeling in organic semiconductors: method and application to two tetracene polymorphs

Patrizio Graziosi[1,*], Raffaele Guido Della Valle[2], Tommaso Salzillo[2], Simone d'Agostino[3], Martina Zangari,[2] Elisabetta Cané[2], Matteo Masino[4], Elisabetta Venuti[2,*]

[1] ISMN – CNR, Consiglio Nazionale delle Ricerche, via Gobetti 101, 40129, Bologna, Italy

[2] Dipartimento di Chimica Industriale "Toso Montanari", Università di Bologna, via Gobetti 85, 410129, Italy

[3] Dipartimento di Chimica "Giacomo Ciamician", Università di Bologna, Via F. Selmi, 2, 40126 Bologna, Italy

[4] Dipartimento di Scienze Chimiche, Della Vita e Della Sostenibilità Ambientale & INSTM-UdR Parma, Parco Area delle Scienze, 17/A, 43124 Parma, Italy

[*] patrizio.graziosi@cnr.it, elisabetta.venuti@unibo.it

## Abstract

We have developed a first-principles method to calculate the electron-phonon coupling for specific modes and *q*-points in the Brillouin Zone for crystalline organic semiconductors. Using the obtained coupling strengths, we propose an approach to compute the temperature-dependent mobilities of electrons and holes.

To validate our treatment of the electronic structures and vibrational properties, we calculate the Raman spectra in the lattice-phonon region and compare them with experimental data. We then compare the computed mobilities with available data for single crystals of naphthalene, anthracene and tetracene (bulk polymorph), finding good agreement within the experimental range, especially when accounting for possible charged impurities. Finally, we discuss the observed differences between tetracene polymorphs.

## I. Introduction

Organic Semiconductors (OSCs) with proper functionalization have emerged as attractive candidates as active layers in several new electronic technologies. These encompass various applications from displays (OLED) to unconventional flexible, stretchable and/or wearable devices, phototransistors, easy processable large area electronics, organic photovoltaics (OPV), energy storage (redox flow batteries, organic electrodes, pseudo-/super-capacitors, gas storage and separation), photocatalytic systems, sensing applications, and others. [1–4] Moreover, OSCs are based on abundant, eco-friendly, and cheap elements, nominally C, O, N, S, with negligible content of critical raw materials. This hallmark is crucial in the development of the next generation of sustainable functional materials.

However, the widespread application of OSCs in electronics-related applications is hampered by the lack of predictive understanding of the inter-relationship between solid-state packing and performance of the device (whatever type of device). This challenge also arises from the existence of multiple crystal structures, known as polymorphism, for a given molecular compound. Each polymorph exhibits a distinct vibrational pattern, leading to variations in electron-phonon coupling (EPC), thereby influencing the transport characteristics of OSCs. [4–6] Indeed, low-frequency vibrational modes, such as translations and rotations of the entire molecule, play a pivotal role in introducing dynamic disorder. This dynamic disorder, in turn, affects on-site energies and electronic transfer integrals, ultimately shaping the macroscale electronic properties of the semiconductor. Therefore, understanding polymorphism is crucial for defining the electronic properties in OSCs at the macroscopic scale.

On the predictive modelling ground, charge transport models are worth a mention. Although sophisticated and computationally intensive approaches exist (such as mean field models, open quantum systems, quantum Monte Carlo, transient localization theory), in most cases approximations leading to analytical expressions are used. [1] These fall into the two categories of hopping and band-like transport via Boltzmann transport equation (BTE). The BTE approach, particularly suitable for orderly packed OSCs, is acknowledged for its precision when accounting for electronic structure details. However, to simplify equations, the effective mass approximation is commonly employed. [7]

In this study, we introduce a methodology for calculating and parameterizing the electron-phonon coupling (EPC) in organic semiconductors, and for using it to assess the OSC mobility. We validate the former by comparing it with experimental vibrational spectra, and the latter with experimental mobility values obtained from single crystals. Our investigation focuses on tetracene, a compound known for its various polymorphs. We observe differences between bulk polymorph 1 and thin-film polymorph, highlighting the need for comprehensive transport measurements on the latter. However, we recognize that the potential coexistence of these two phases, a phenomenon often detected in acenes, may pose a challenge in this task. [8,9]

## II. Computational method

### a. Electronic and Vibrational properties

On this ground, we follow the most effective state-of-the-art approach, [10–12] employing PBE pseudopotentials and the D3-BJ Grimme with Becke-Johnson damping functions for a posteriori VdW correction, [13] *at any stage* of a DFT calculation with VASP. [14,15] We start on the experimental unit cell parameters, optimize the k-points grid using a cutoff energy of 400 eV. After having detected a converging *k*-grid of 5×3×2 and 7×5×3 for P1 and PF polymorphs, respectively, we optimize the energy cutoff for the wavefunctions. Finally, we relax the atomic positions keeping constant the unit cell parameters at the experimental values. Following relaxation, Phonopy packages [16,17] are used to compute and diagonalize the dynamical matrix using a 2×2×2 super-cell – a setup validated to ensure convergence in phonon frequency and 3D density of states (DOS) without any negative frequency, except for the acoustic branches in Γ, for which negative frequency around $10^{-2}$ - $10^{-3}$ THz may be obtained and are considered acceptable. Finally, the Raman spectrum is simulated by using the vasp_raman.py code. [18] When comparing with experimental Raman spectra, the computed intensities have been adjusted by considering excitation wavelength and temperature dependence. [19] We consider this step mandatory to ascertain the suitability of the overall DFT scheme for the system under investigation. To compute the electronic dispersions, after the calculation of the frequencies, which ensures the structure has reached the energy minimum, we repeated the self-consistent calculation on a 7×5×3 *k*-space grid, for each polymorph, then we performed the required not-self-consistent calculations for computing the bands along certain paths, [20] by using 20 steps between each pair of high-symmetry points, for both polymorphs.

### b. Electron-phonon coupling (EPC)

Following the previous procedure, we employ Phonopy to modulate the unit cell along the selected eigenmodes for Γ point and all the relevant high-symmetry ***q***-points within the BZ, as given in the Bilbao crystallographic data center. [21,22] Thus, for each considered ***q*** point and $\nu$ phonon branch, a unit cell distorted along the eigenmode vibrational coordinate is generated. Phonopy code automatically weights for the atomic mass in this process. In the tight-binding approximation the dispersion energy can be approximated along a 1D direction to ~ $2\, t_W\, cos(\boldsymbol{k} \cdot \boldsymbol{a})$, and that leads to a bandwidth $B_W$ related to the transfer integral $t_W$ by the relation $B_W = 4t_W$. [23] As a result, the comprehensive bandwidth across the full 3D Brillouin Zone, and hence the corresponding transfer integral, can be obtained from a first principle DFT electronic structure calculation.

Under this conceptual framework, and taking into account the traditional approach to EPC [24,25] and the conventional definition of Deformation Potentials, [26,27] the EPC constant for each $q$-point and phonon branch $\nu$ can be defined as :

$$D_{q,v}^{n,n} = \frac{\partial t_W^n}{\partial r_{q,v}} \qquad (1)$$

$$D_{q,v}^{n,m} = \frac{\partial \Delta_s^{n,m}}{\partial r_{q,v}} \qquad (2)$$

In Eq. s (1)-(2), $n$ and $m$ are the band indexes, $t_W^n$ is the bandwidth of the band of index $n$, $\Delta_s^{n,m}$ is the Davydov splitting between the bands $n$ and $m$, computed as the energy difference between the energy barycenters of the bands, $r_{q,v}$ is the average displacements of all the atoms in the structure distorted along the eigenmode $(q, v)$. Thus, Eq. (1) is related to *intra*-band processes, i.e. will be used when the carrier scattering involves initial and final states in the same band, while Eq. (2) is related to *inter*-band processes.

Operatively, after a self-consistent DFT calculation for each $(q,v)$ modulated/eigen-distorted structure, we perform a non-self-consistent calculation and save the obtained electronic structure in .bxsf format [28] with the c2x code. [29] Ad hoc routines were developed to extract the EPC parameters as in Eq. s (1) and (2). Finally, since we will be considering states, and transitions, in the whole 3D electronic structure, a DOS-weighted average of the computed EPC, across the $q$-points for each branch $v$, was performed to extract band-index specific EPC parameters to be used throughout the whole BZ:

$$D_{n,m,v} = \frac{\sum_q D_{q,v}^{n,m\,2} DOS_{q,v}}{\sum_q DOS_{q,v}} \qquad (3).$$

The selection of the vibrational modes to be considered in this protocol relies on the phonon DOS specifications and is made considering the modes which are bundled in the DOS. For tetracene, this encompasses the 16 modes of lowest frequency, with the additional eventual inclusion also of modes 17-20. This approach is exemplified in section III. The comprehensive EPC parametrized in Eq. (3) are regarded as the proper deformation potentials for inelastic processes involving non-polar phonons in the mobility calculation, as detailed in the next sub-section.

The deformation potential computed as in Eq. s (1) and (2) compares with the commonly employed electron-phonon interaction matrix element $|g_{m,n,q,v}|$ definitions as follows: [30,31]

$$|g_{m,n,q,v}|^2 = \frac{\hbar}{2M\omega_{q,v}} D_{q,v}^{n,m\,2} \qquad (4),$$

where *n* and *m* are the band indexes, *q* and *v* are the phonon wave-vector and branch index, *M* is the mass of the unit cell and $D_{q,v}^{n,m}$ the EPC strength defined in Eq. s (1) and (2). We use the equation above also to extract the deformation potential for the *ElecTra* code from the Electron-Phonon Average (EPA) approximation calculation, performed with the epa.x program inside the Quantum Espresso suite. [32] The phonon energy was chosen to be the corresponding energy difference between initial and final states in the epa.x output. The e-ph matrix values corresponding to the same phonon energy were average with the wight and multiplicity reported in the epa.x output.

Moreover, we evaluated a deformation potential for the elastic scattering with acoustic phonons near Γ, Acoustic Deformation Potential scattering mechanism (ADP), by computing the variation of the bandwidth (intra-band process) or Davydov split (inter-band process), with the same philosophy of Eq. s (1) and (2), in respect of the relative cell volume variation. [26,27,33]

$$D_{ADP}^{n,n} = <\frac{\partial t_W^n}{\frac{1}{3}\left(\frac{\Delta V_c}{V_c}\right)}> \quad (5)$$

$$D_{ADP}^{n,m} = <\frac{\partial \Delta_S^{n,m}}{\frac{1}{3}\left(\frac{\Delta V_c}{V_c}\right)}> \quad (6).$$

We reduced and expanded the cell volume $V_c$ by increasing/decreasing the cell axes by a 1/1000 factor, and then averaged between compression and dilation. The sound speed, required to evaluate the ADP scattering rate, was taken from [34].

### c. Mobility

Crystalline OSCs exhibit a well-defined crystal structure, which manifests in a distinctive vibrational fingerprint. Adopting a description based on the electronic dispersions in the BZ of the reciprocal lattice [35] the mobility *μ* is evaluated from the conductivity *σ* as:

$$\mu_{ij(E_F,T)} = \frac{\sigma_{ij(E_F,T)}}{n \cdot q_0} \quad (7)$$

where *i* and *j* are the Cartesian components *x*, *y*, and *z*, of the mobility and conductivity tensors, $E_F$ is the Fermi level, *T* the temperature, *n* the carrier density and $q_0$ the electronic charge. The conductivity is computed in the context of the linearized BTE as:

$$\sigma_{ij(E_F,T)} = q_0^2 \int_E \Xi_{ij}(E) \left(-\frac{\partial f_0}{\partial E}\right) dE \quad (8)$$

The integrand of Eq. (8) contains the Transport Distribution Function (TDF) $\Xi_{ij}$ and the energy derivative of the equilibrium Fermi-Dirac distribution $f_0$. The TDF is defined as:

$$\Xi_{ij}(E) = \frac{2}{(2\pi)^3} \sum_n \sum_{k_{n,E}} v_{i,k_{n,E}} v_{j,k_{n,E}} \tau_{i,k_{n,E}} g_{k_{n,E}} \quad (9)$$

In Eq. (9) $v$ is the band velocity, $\tau$ the relaxation time, and $g$ the electronic DOS. All these quantities are specific of each individual transport state $k_{n,E}$, where $k$ is the wave-vector, $n$ indicates the band index and $E$ its energy. So, the sum runs over all the transport states identified by their momentum, belonging to all the bands, having a certain energy. The DOS $g_{k_{n,E}}$ is defined as $\frac{dA_{k_{n,E}}}{|\vec{v}_{k_{n,E}}|}$, where $dA_{k_{n,E}}$ represents the area of the surface element of the constant energy surface to which the $k_{n,E}$ state belongs, associated to each specific $k_{n,E}$ state. Therefore, the sum in Eq. 6 is performed for each constant energy surface to compute the energy dependent TDF which will be then integrated as in Eq. (8). In this work, the tetrahedron method has been employed to construct the constant energy surfaces and extract the related quantities. [36,37]

The relaxation time of the state $k_{n,E}$ for the transport along the direction $i$, related to scattering with a phonon belonging to the branch $\nu$, is evaluated from the inelastic scattering with non-polar phonons as (we omit the index $\nu$ for clarity): [26,27,37–39]

$$\frac{1}{\tau_{i,k_{n,E}}} = \frac{1}{(2\pi)^3}\sum_{k'}\frac{\pi D_{n,n'}^2}{\rho\omega_0}\left(N_{\omega_0} + \frac{1}{2}\mp\frac{1}{2}\right)g_{k'_{n',E'}}\left(1 - \frac{v_{i,k'_{n',E'}}}{v_{i,k_{n,E}}}\right) \quad (10)$$

where $D_{n,n'}$ is the deformation potential related to the electron-phonon scattering between the initial band $n$ and the final band $n'$, which includes intra- and inter-band processes, evaluated from Eq. (1) or (2), respectively. $\rho$ is the mass density, $\omega_{0,\nu} = \frac{\sum_{\omega_\nu}\omega_\nu \cdot DOS(\omega_\nu)}{\sum_{\omega_\nu}DOS(\omega_\nu)}$ is the effective frequency for the branch $\nu$ evaluated from a DOS-weighted average over the selected portion of the phonon spectrum, $N_{\omega_0}$ represents the phonon Bose-Einstein statistical distribution, $g_{k'_{n',E'}}$ is the DOS of the final state, belonging to the band $n'$ and at energy $E'$, which is either increased or decreased by $\hbar\omega$ for absorption or emission processes, respectively, denoted by " – " and " + " signs. The term $\left(1 - \frac{v_{i,k'_{n',E'}}}{v_{i,k_{n,E}}}\right)$ approximates the momentum relaxation time, [38–40] which is the relevant type of relaxation time for computing transport coefficients. [27] Note that this definition of relaxation time bears similarity to what appears in other works on OSCs. [41] The calculation of the mobility is performed using the *ElecTra* simulator. [35,42]

To perform the charge transport calculations, an additional non-self-consistent calculation on a finer mesh is needed. In this study, we adopted a 30 × 20 × 15 mesh with cutoff energy of 900 eV for the Polymorph 1 (P1) and 28 × 20 × 12 mesh with cutoff energy of 1000 eV for the thin film polymorph (PF), reaching the limit of the available node memory of the computing cluster. However, tests were run from a 7 × 5 × 3 grid onward and we found that a 25 × 15 × 10 grid provides convergence. Most important, the energy resolution used to construct the constant energy surfaces is set at 1 meV, meaning that each surface is calculated every 1 meV for each band, from the band edge up to ~ 0.4 eV. Such fine resolution proves to be essential for the treatment of the flatter bands of the OSC compared to inorganic compounds. [37,43–45] Due to the large bandgap of tetracene, we perform unipolar calculations separately for electron mobility in the conduction band (CB) and hole mobility in the valence band (VB).

In addition to the main quantities described above, this work includes other quantities, computed when other processes needed to be accounted for. In the investigation of the role of punctual charged impurities, such as acceptors in the *p*-type case, the scattering with carriers was modeled as Ionized Impurity Scattering (IIS) within the Brooks-Herring framework:

$$\frac{1}{\tau^{IIS}_{i,k_{n,E}}} = \frac{1}{(2\pi)^3}\sum_{k'}\frac{2\pi}{\hbar}\frac{Z^2 e^4}{k_s^2 \varepsilon_0^2}\frac{N_{imp}}{\left(|k-k'|^2+\frac{1}{L_D^2}\right)^2} g_{k'_{n',E'}}\left(1-\frac{v_{i,k'_{n',E'}}}{v_{i,k_{n,E}}}\right) \quad (11)$$

where $Z$ is the impurity charge, $k_S$ the relative static permittivity, and $L_D$ the screening length computed as

$$L_D = \sqrt{\frac{k_S \varepsilon_0}{q_0}\left(\frac{\partial n}{\partial E_F}\right)^{-1}} \quad (12).$$

We also discuss a diffusion mobility, defined from the Einstein relation as

$$\mu^D_{ij(E_F,T)} = \frac{q_0 D_{ij(E_F,T)}}{k_B T} \quad (13)$$

where the diffusion coefficient $D_{i,j(E_F,T)}$ is computed starting from a DOS-weighted average of the transport state $v_{i,k_{n,E}} v_{j,k_{n,E}} \tau_{i,k_{n,E}}$, for each band, summing over all the bands:

$$D_{ij(E)} = \sum_n \frac{\sum_{k_{n,E}} v_{i,k_{n,E}} v_{j,k_{n,E}} g_{k_{n,E}}}{\sum_{k_{n,E}} g_{k_{n,E}}} \quad (14)$$

An average over the carrier energy, also weighted by $\frac{\partial f_0}{\partial E}$, which approximates to $f_0(1-f_0)$, is performed to account for the states involved in transport:

$$D_{ij(E_F,T)} = \frac{\int D_{ij(E,E_F,T)} g_{(E)} \frac{\partial f_0}{\partial E} dE}{\int g_{(E)} \frac{\partial f_0}{\partial E} dE} \quad (15).$$

Finally, in the mean free path $\lambda$ evaluation, a DOS-weighted average of the energy dependent mean free path is computed as:

$$\lambda_{i(E_F,T)} = \frac{\sum_E \lambda_{i(E,E_F,T)} g_{(E)}}{\sum_E g_{(E)}} \quad (16)$$

where $i$ is the Cartesian coordinate, $g_{(E)}$ the total electronic DOS, and each $\lambda_{i(E,E_F,T)}$ term has the expression:

$$\lambda_{i(E,E_F,T)} = \frac{\sum_{n,k_{n,E}} |v_{i(k,n,E)} \tau_{i(k,n,E,E_F,T)}| g_{k_{n,E}}}{\sum_{n,k_{n,E}} g_{k_{n,E}}} \tag{17}$$

### III. Results and Discussion

#### a. Electronic Structure

The electronic dispersions of the two investigated tetracene polymorphs are reported in Figure 1, where in red and blue are depicted the valence and conduction bands, respectively. The paths have been chosen according to [46]. We highlight with a green circle the valence band maxima (VBM) and the conduction band minima (CBM). For P1, the CBM appears to be at the R point while the VBM is located at the V point. For PF, the CBM happens at X whereas the VBM is along the Γ–U line.

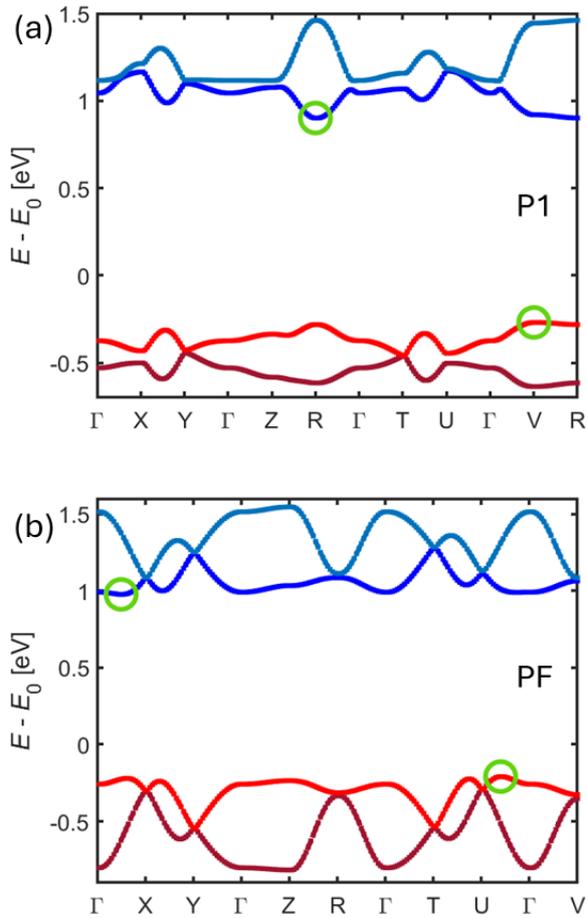

Figure 1: electronic band dispersions along paths between high symmetry points in the reciprocal unit cell for P1 (a) and PF (b). Red and blue colors indicate valence and conduction bands, respectively, while the green circles highlight the bands extrema.

The electron 3D DOS and band velocity, computed in the reciprocal unit cell on the mesh used for the subsequent transport calculations and over the adopted energy range, are depicted in Figure 2. The two VBs and two CBs lie quite separated in P1, when the DOS drops and then rises again. In contrast, these bands are more bundled in PF, leading to higher total band velocity in this polymorph.

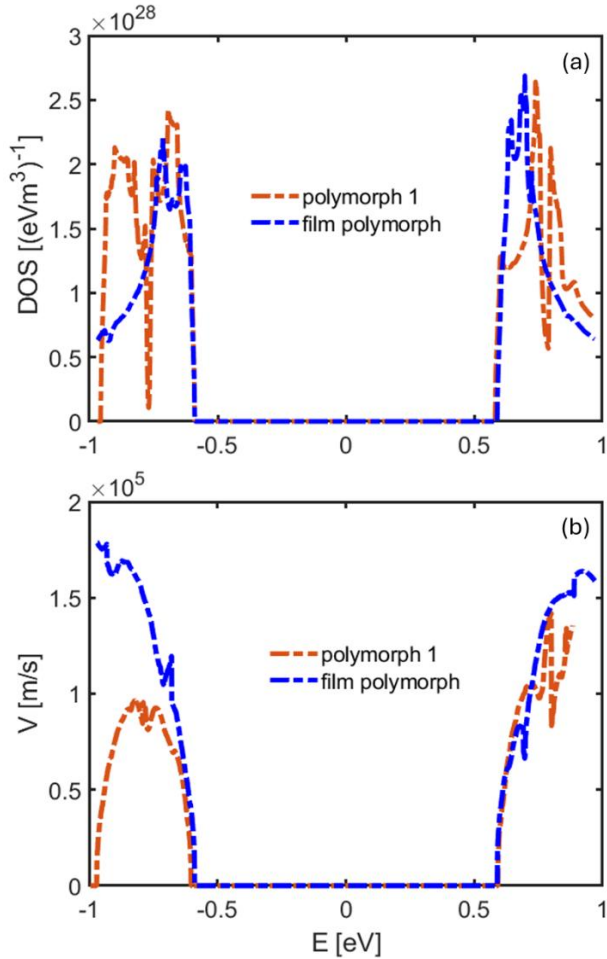

Figure 2: (a) density-of-states (DOS) and (b) band velocity, in S.I. units, for the two tetracene polymorphs investigated. Two CB bands and two VB bands can be distinguished in the P1 where the DOS falls down and rises again, so that bands have little energy overlap; in the PF the overlaps are more significant. The computed bandgaps are narrower than the experimental ones, as common in DFT calculations, [47–49] however this does not affect the computed unipolar mobility.

### b. Vibrational properties

The vibrational properties were computed for P1 [50] and PF [51,52] phases. The low temperature Polymorph II [53,54] was not considered as the atomic coordinates of this structure are, to the best of our knowledge, unknown. The phonon dispersions and the corresponding DOS are reported in Figure 3. It is important to note that the first 16 modes are bundled together throughout the BZ. This feature is highlighted in Figure 3a, where the DOS for the two investigated polymorphs are reported. The dispersions along directions connecting high symmetry $q$-points are shown in Figure 3b and 3c for P1 and PF, respectively. It is possible to observe how the 16 modes below 4.5 THz cross and intersect each other in the BZ. Additionally, in the thin film case, two upper modes slightly overlap with the lower 16; however, this minor overlap will be neglected in the following analysis. Since the lower 16 modes are bundled, we consider them as 16 scattering channels, following the common approach in semiconductors, [27] and due to the dispersion through the BZ, we assign to each of them an effective phonon frequency $\omega_{0,\nu}$, computed as a DOS-weighted average of the phonon frequencies of that mode, as described in section IIc. Because of their boundles character, we consider all of them and cut our region of interest at the first drop to zero of the phonon DOS:

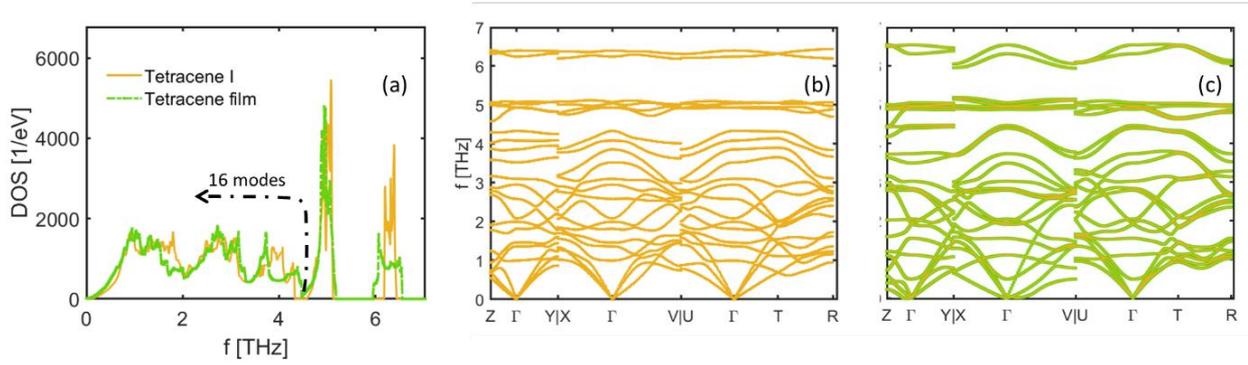

Figure 3: phonons DOS (a) and dispersions (b-c) for the P1 (orange) and the PF (green) polymorphs, respectively.

We further investigated the character of these modes by analyzing their Participation Ratio (*PR*) defined as: [55,56]

$$PR_{q,\nu} = \frac{1}{N_a} \frac{\left(\sum_a \frac{|e_{a,q,\nu}|^2}{m_a}\right)^2}{\sum_a \left(\frac{|e_{a,q,\nu}|^2}{m_a}\right)^2} \quad (18)$$

Where $e_{a,q,\nu}$ is the phonon eigenvector for mode $\nu$ in point $q$ and atom $a$, $m_a$ is the mass of atom $a$, $N_a$ is the total number of atoms in the unit cell and the sums run over all the atoms in cell.

This parameter provides a normalized estimation of the fraction of atoms involved in the $\nu$-th vibrational mode. The spatial extension of such a group is linked to the localized or extended nature of that mode: for extended modes $PR \sim 1$, while localized modes have a smaller ratio, down to the limit $PR = 1/N_a$ when the mode involves only a single atom.

The *PR* values for the first 16 modes are shown in Figure 4, for P1 and PF in (a) and (b), respectively. We can notice that at the zone boundaries, many modes display also a significant localized character. The extended character of the modes is more distributed over the BZ for the PF, with high *PR* values also at the zone boundaries, for some modes and $q$-points. When averaging the *PR* over this range, the thin-film phase shows a value nearly 3 % higher, 0.55 vs 0.53. Focusing on the nominal intermolecular modes, specifically the lowest 12 modes, the thin-film phase has a *PR* ~ 5 % higher, 0.61 vs 0.58. A detailed description of the nature of rigid molecule modes at Γ can be obtained by projecting the eigenvectors along the rotational and translational coordinates in the molecular inertia axes system. In Figures 4c and 4d, the analysis of such decomposition of the lowest 16 modes in Γ is reported.

DFT calculations can be validated by comparison with spectroscopic experiments, which probe frequencies at the Γ point. To accomplish this task, the off-resonant Raman activities were computed with the vasp_Raman.py program. [18] This program uses VASP as back-end to compute the polarizability with the approach of the finite displacements, and returns the Raman activity of the selected modes. The results of the calculations in the low frequency range for P1 are compared with the experimental spectrum [53] in Figure 5a, with a zoom in the range of the experimental data in Figure 5b. The spectra are drawn as Lorenzian bands with a fwhm of 1/3 of the mean distance between the frequencies, chosen to conform to the experimental features of P1. The positions of the peaks agree within a very few cm$^{-1}$, less than 5, reported also in the Supplemental Material, [57] thus confirming the validity of our description in terms of pseudopotentials and electronic structure. [11,12,58,59] The experimental relative intensities map satisfactorily on the simulated ones once the laser excitation frequency and the measurement temperature of the experiments are accounted for. [19] This is performed by using the formula $I = I_0 \frac{\nu}{(\nu-\nu_0)^4}\left(1 - \exp\left(\frac{-h\nu}{kT_0}\right)\right)$, where $I_0$, $\nu_0$, $T_0$, and $\nu$ are the measured Intensity, excitation frequency, temperature and vibration frequency, respectively, $h$ is the Planck constant and $k$ the Boltzmann constant. $I$ is the adjusted experimental intensity.

The fact that the experimental and computed peak positions agrees generally better than intensity ratios is not surprising, as it may stem on the unknown crystal orientation of the measured anisotropic specimen, [10,59] which determines which components of the polarizability matrix, and thus the band intensities, were actually probed in the experiment. Additionally, figures 5c and 5d present the Raman spectrum of the polymorph, not yet reported experimentally, highlighting clear differences that can guide future experimental spectroscopic studies. [12,58,59]

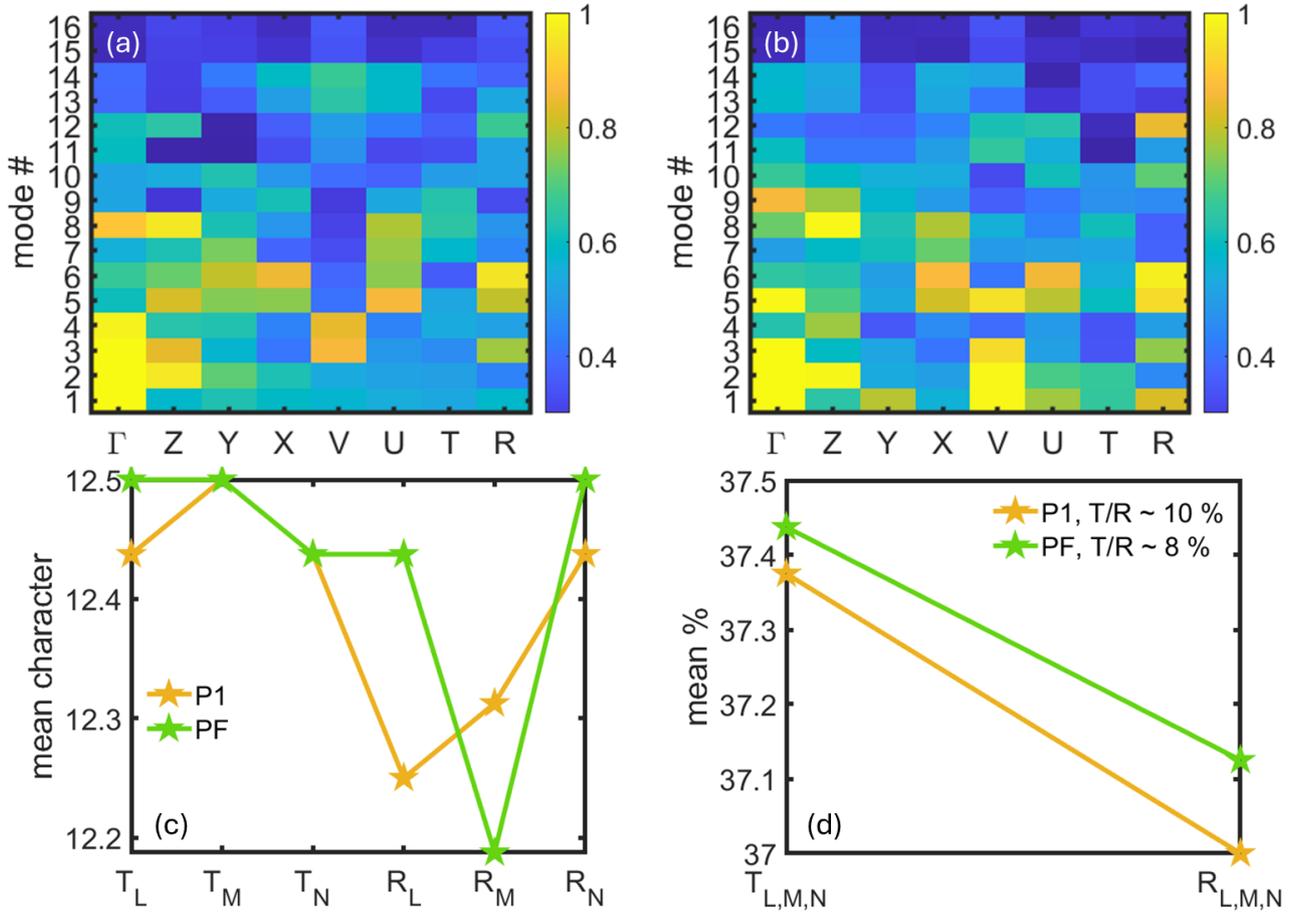

Figure 4: *PR* values at different BZ high symmetry ***q***-points, for the lower 16 modes, (a) P1, (b) PF. We observe a slightly higher extended character in the PF modes, especially in the lower modes and far off Γ, whereas in the P1 the high extended character is mostly limited to BZ center. (c) Mode decomposition at Γ along translational (T) and rotational (R) eigenvectors, averaged for all the modes. (d) The mode average L, M, and N components of T and R have been summed, it appears that the PF modes have an average larger extended character, and a larger R character in comparison to the T one.

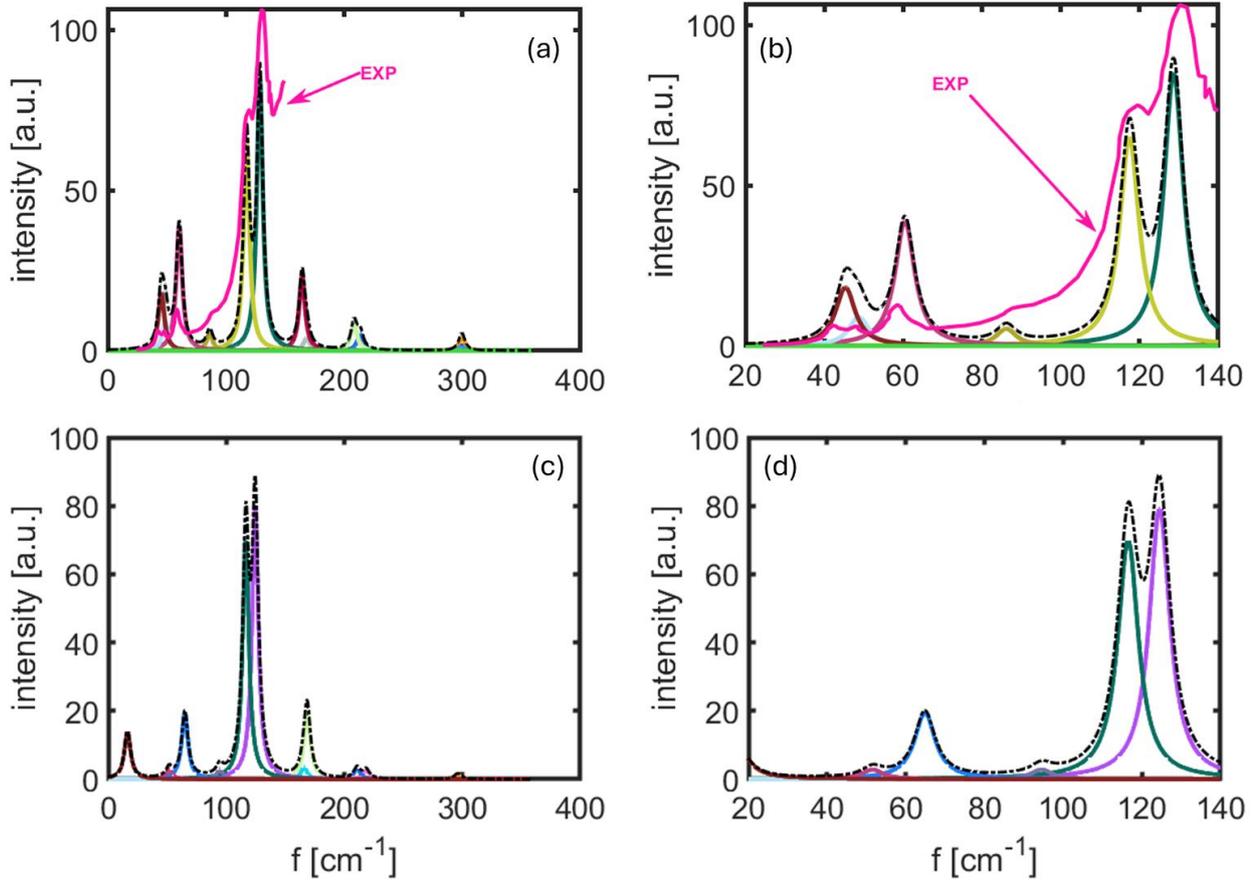

Figure 5: (a) and (b), validation of the DFT protocol that compute the electronic and vibrational structure via comparison between computed and experimental Raman spectra of the P1, the experimental data, in magenta line, have been obtained from ref. [53]. The experimental intensities have been transformed into Raman activities by accounting for measurement temperature and laser frequency. (c) and (d), the Raman spectra of the PF are shown, on the same frequency ranges.

### c. EPC parameters

In this section, we present the EPCs evaluation according to Eq. s (1)-(3). They have been computed with Phonopy for unit cells with atoms displaced along each eigenmode, employing an input amplitude tag of 0.1, which results in an average displacement of the order of $10^{-4}$Å. The test on the convergence of such values with respect to the displacement is detailed in the Supplemental Material. [57] Figures 6 and 7 show the EPCs computed for P1 and PF, respectively, from the different bands and the Davydov splitting. The terms CB and VB refer to conduction band and valence band, respectively, with $1^{st}$ and $2^{nd}$ indicating the order from the edge, with the former having the lower minimum. In Figures 6 and 7 the horizontal axes span the high symmetry BZ points and the mode number, while the z-axes represent the magnitude of the EPC. Note that the colors are just for representative purpose, there is no color scale as the EPC strength is represented by the value on the vertical axis.

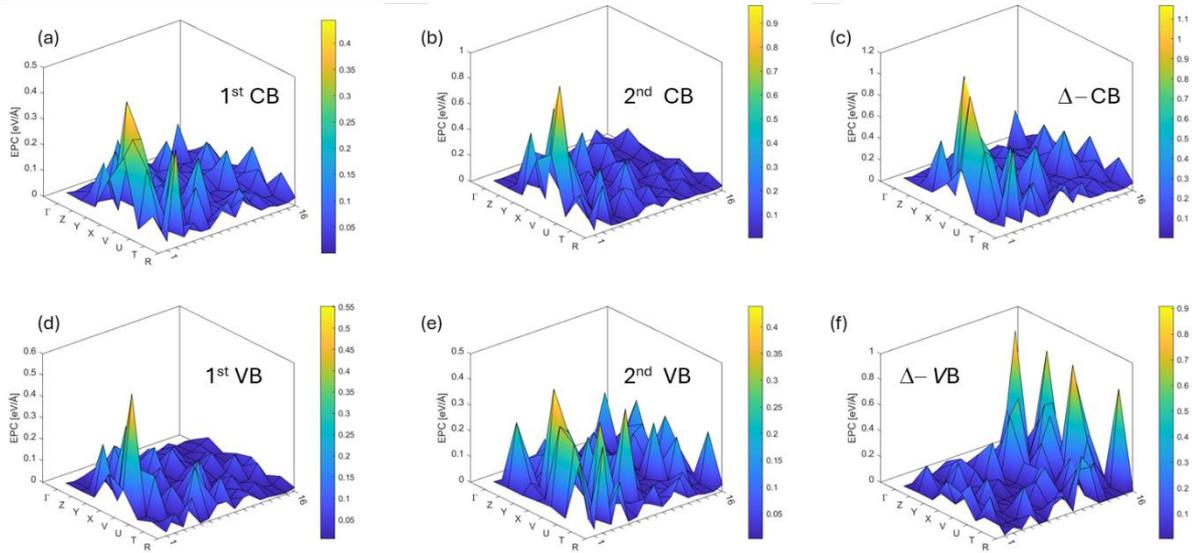

Figure 6: EPC values in eV/ Å for tetracene bulk, comprising the lowest 16 modes at 8 high symmetry $q$-points in the BZ, as indicated. (a) and (b), EPC computed with Eq. (1) for the two CBs, (c) EPC computed from the Davydov splitting as in Eq. (2). (d)-(f), same as (a)-(c) for the VBs.

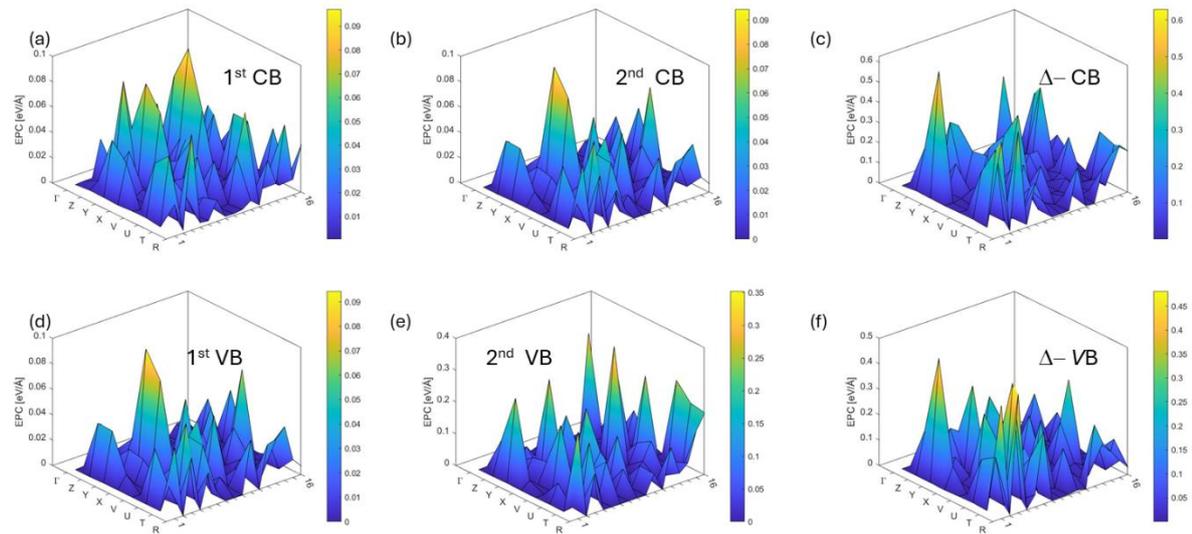

Figure 7: EPC values in eV/ Å for tetracene PF, comprising the lowest 16 modes at 8 high symmetry $q$-points in the BZ, as indicated. (a) and (b), EPC computed with Eq. (1) for the two CBs, (c) EPC computed from the Davydov splitting as in Eq. (2). (d)-(f), same as (a)-(c) for the VBs.

From Figures 6 and 7 we can deduce that certain modes, at specific *q*-points of the BZ, i.e. with a given momentum, lead to very strong EPC, whereas at other *q*-points the EPC strength is much lower. Importantly, the stronger EPC values are found not to be at the Γ point. Since our task is to describe the scattering across the entire BZ, in every direction, we must average them. [44] This holds not only for each individual mode, but also for all of them together, given their strong bundled feature. Thus, the values in Figures 3 and 4 must be weighed for the DOS and averaged as from Eq. (3). Mode and *q*-points specific EPC values are listed in the Supplemental Material, part B. [57]

The EPC values obtained through this procedure are reported in the Supplemental Material, [57] whereas in Table I we provide the sum of the EPC of each mode as a one-number rough indication of the EPC strength, labelled EPC*, in units of eV/Å. The same Table also lists the band widths (BWs), which are related to the transfer integrals between translationally equivalent molecules (α or β) in different unit cells. [60] Thus, a larger BW indicates a stronger interaction between translationally equivalent molecules in different unit cells of the crystal. This might lead to higher insensitivity to the modulation carried by the vibrational modes.

|  | VB1 | VB2 | inter-band | CB1 | CB2 | inter-band |
|---|---|---|---|---|---|---|
| P1 – EPC* | 1.27 | 1.69 | 2.80 | 1.54 | 2.35 | 3.39 |
| P1 - BW | 0.191 | 0.189 | 0.013 | 0.268 | 0.351 | 0.060 |
| PF – EPC* | 0.32 | 1.36 | 1.88 | 0.41 | 1.0 | 2.17 |
| PF - BW | 0.525 | 0.342 | 0.252 | 0.30 | 0.467 | 0.198 |

Table I: EPC, in eV/Å, and bandwidth BW, in eV, for the CB and VB of the two polymorphs. The column "inter-band" contains, for the EPC, the inter-band scattering term, for the BW, the amount of the overlap between the two bands. It is possible to observe that the PF has much broader bandwidths.

### d. Mobility

In this section, the mobility is computed using the approach provided in section II.c. The mobility tensor derived using equations (7)-(10) is in Cartesian coordinates.

First, we project the electric field $\vec{\mathcal{E}_l}$ along the crystallographic axes. This is achieved by inverting the lattice vector matrix, expressed in the POSCAR VASP file:

$$\vec{\mathcal{E}_l} = l * \text{inv}(A) \tag{19}$$

where *l* is the intended direction in the internal coordinates and *A* is the lattice vector matrix. This allows us to express the electric fields along the internal cell axes, $\vec{\mathcal{E}_a}, \vec{\mathcal{E}_b}$, and $\vec{\mathcal{E}_c}$ in Cartesian coordinates. Next, from the conductivity tensor $\bar{\bar{\sigma}}$, given by Eq. (8), we compute a current density as:

$$\vec{J_l} = \bar{\bar{\sigma}}\vec{\mathcal{E}_l} \tag{20}.$$

Thus, we obtain a component of the conductivity tensor in internal coordinates:

$$\sigma_l = |\vec{J_l}|/|\vec{\mathcal{E}_l}| \qquad (21)$$

and the related mobility:

$$\mu_l = \sigma_l/n \cdot q_0 \qquad (22).$$

Using this approach, we can link the computed mobility tensor in Cartesian coordinates to the mobility measurable along specific crystal directions.

We first aim at validating our approach by comparing computed and experimental data. Considering that the reported experimental values vary widely, mostly due to extrinsic phenomena, mobilities measured in single crystals should be regarded as the relevant reference. [61,62]

For the sake of genericity, we apply the proposed method not only to the model polymorphic tetracene system, [62,63] but also to other oligoacenes for which time-of-flight (ToF) measurements [64] of mobility on high quality single-crystal are available, namely naphthalene, [65] anthracene. [66]

These comparisons are reported in Figure 8a, 8c, and 8d, for tetracene, naphthalene, and anthracene, respectively. The experimental mobilities are plotted in black, and the computed mobilities are given with different colors as in the legends, given by assuming the Fermi level placed at around ¼ of the bandgap. For naphthalene and anthracene, we report the mobilities $\mu_a$ and $\mu_b$ along the *a* and *b* axes, respectively, as well as $\mu_{\perp_{ab}}$ along the direction perpendicular to the *ab* plane, also called *c\** or *c'*. [65,66] For tetracene, we report both the experimental minimum mobility, measured in tetracene single-crystal OFETs, [62] and the ToF mobility at 300 K. [63] For the cases of naphthalene and tetracene, we report also the mobility computed considering the elastic scattering with acoustic phonons near Γ (ADP), in gray, demonstrating that its contribution is negligible. We specify that the ADP deformation potentials computed with Eq. s (5) and (6), given in part D of the Supplemental Material, [57] are lower than the ones previously reported in literature with the method of the relative displacements of the band edges. [67] For naphthalene, Figure 8c, we also report in red the mobility along the *b* axis computed with the EPA method, [32] which is based on Density Functional Perturbation Theory (DFPT) technique. The severe underestimation of the mobility computed within the EPA approximation is likely due to the coarse grid we used, see table II. However, we chose the computational settings to fit in with the 24 hours wall-time limit of the HPC infrastructure we used.

We observe that the computed mobilities are of the correct order of magnitude, though somewhat larger, which is expected considering that we are assuming ideal crystals. The fact that in the case of anthracene, the computed $\mu_b$ is lower than the experimental values, is not expected. However, as we show in part E of the Supplemental Material, [57] the mobility of anthracene is dominated by the inter-band scattering. It is possible that our proposed Eq. (2) for the inter-band scattering, based on the Davydov splitting, requires further corrections; for example, a decrease of around 25% of the considered split brings the computed values above the experimental ones.

The mean free path $\lambda$ estimated with Eq. (16) at $T = 300$ K for the investigated systems, considering the Fermi level at around ¼ of the bandgap, are reported in Table II, in units of Å. We observe that they reflect the anisotropy of the mobility. For naphthalene, they are very close to the values extracted from experiments, [60] and, except for the tetracene PF phase, they are generally in the order of the inter-molecular spacing, as expected. [63]

|  | $\lambda_a$ | $\lambda_b$ | $\lambda_c$ |
|---|---|---|---|
| tetracene P1 | 2.04 | 2.61 | 0.91 |
| tetracene PF | 28.3 | 35.63 | 5.78 |
| naphthalene | 1.58 | 2.80 | 2.0 |
| anthracene | 1.64 | 2.36 | 1.77 |

Table II: mean free path $\lambda$ along the three crystal axes for the tetracene polymorphs and the other two model systems, values in Å.

We highlight that the experimental single crystal mobility appears to be limited by the traps at the metal/OSC contact interface and that shallow traps in this region are likely to strongly decrease the mobility by more than 70 %. [62] Based on this, we included punctual defects in the model. These defects act as electron acceptors and behave like *p*-dopants. We also assume that the carrier density is completely due to these impurities, which consistently downshift the Fermi level towards the band edge. From the perspective of carrier scattering, the defects are treated in the framework of the IIS described in section II.c. Their charge was set either equal to one or ½, at densities comparable to those estimated for OSC crystals. [68,69] The choice of a ½ charge should simulate a situation where the charge of the ionized acceptor is effectively spread over more molecules, resembling an extended charged defect. The results for two densities of ionized defects are plotted in magenta in Figure 8a. We can see that the calculated mobility values fall within the experimental range, although with a reduced slope in the T dependence. This demonstrates that mobilities can be brought within the interval of the experimental values by considering defects, although, undoubtedly, a better description of these would be necessary. This is even more true for the case of FET mobility, where the determinant role of scattering at the boundaries between differently oriented regions has been observed by in-operando kelvin-force measurements. [70] This scattering is likely responsible of a measured tetracene mobility [71,72] which is lower than the one estimated for bulk single crystals. [62]

Additional insights are gained from the analysis of the power law exponent in $\mu \sim T^n$. By fitting all the experimental data together, black line in Figure 8b, an exponent of -2.56 is computed. When splitting the analysis to distinguish between high and low temperature regimes, the low temperature one (red dotted line) gives an exponent of -4.50, while the low temperature (green dotted line) gives an exponent of -1.77. The computed BTE mobility (orange stars and orange dash-dot lines) returns an exponent of -1.69, which is very close to the experimental value at low $T$, whereas for the diffusion mobility $\mu_D$, Eq. s (13)-(15), this exponent is much lower, -0.66. This convinced us that the description based on the DFT electronic structure, which accurately reproduces the experimental Raman spectra, can be regarded as a solid basis also for transport simulations in crystalline OSCs. Thus, our findings suggest that the common BTE mobility is more suitable than the diffusion mobility to describe the OSCs. We note that when considering punctual ionized impurities in the BTE, magenta lines in Figure 8a, $n$ falls between -0.90 and -1.07, calling for improved models for defects in OSCs. Finally, we performed a mobility calculation that included, at the same foot of the lowest 16 modes, the next 4 modes between 4.5 and 5.5 THz, which form a bundled group in the DOS right above 4.5 THz. However, no significant differences were detected in the mobility values despite this additional scattering process involving a larger ensemble of modes. This suggests that our approach, that considers the lower bundled modes up to the first drop in the DOS, together with the method to compute and parametrize EPC, is suitable to describe the mobility of the OSC under investigation.

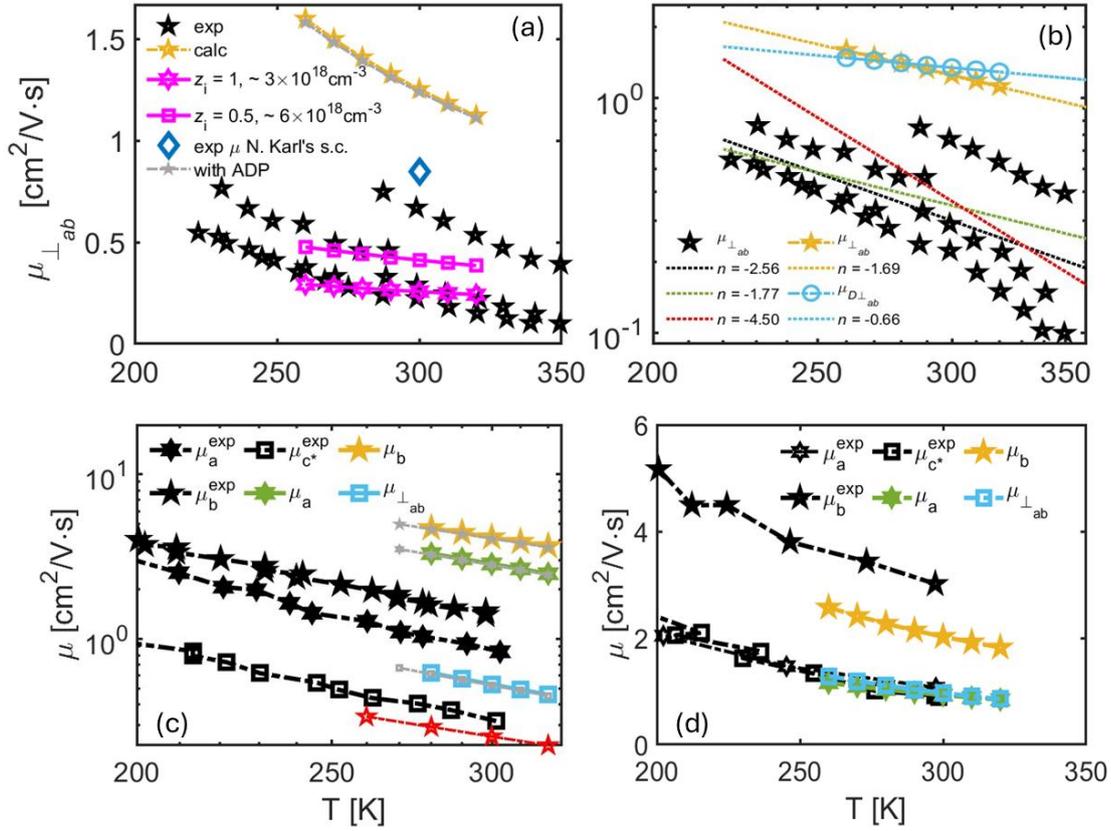

Figure 8: validation of the proposed approach with respect to the experimental hole mobility values for tetracene, in (a) and (b), naphthalene in (c), and anthracene in (b), along different crystal direction as denoted. [62,63,65,66] For the case of tetracene, the mobilities computed with the addition of punctual ionized defects, considered as electron acceptors, at two different concentrations and ionized charges, are reported in (b). The fractional charge aims at mimicking extended charged defects. By including these, the match between the experimental and computed mobilities improves, but at the cost of a change in the $T$ dependence. However, this suggests that the overestimation of the computed mobility is likely due to the neglect of extrinsic charged defects. The diffusion mobility computed from Eq. s (13)-(15) is shown in (b), indicated as $\mu^D_{\perp_{ab}}$. The corresponding fits are displayed, as explained in the main text. In (c), for naphthalene, we add the mobility computed with the EPA method.

Eq. (19) allows us to express the electric field in the internal coordinates, enabling the analysis of the mobility anisotropy within the crystal planes. This is demonstrated for the hole and the electron transport in P1 in Figure 9a and 9b, respectively. For each plane, the lines are plotted by using over $10^6$ random directions on the plane. A uniaxial anisotropy appears, that means that a direction of higher mobility is present, which is nearly parallel to the $b$ axis. This direction lies in the $ab$ plane, which is generally the contact plane for OFETs, and is consistent with what observed in other OSCs like pentacene or rubrene. [73,74] Moreover, we obtain an anisotropy ratio of the mobility in the $ab$ plane of 3, in good agreement with the experimental findings, which is around 2. [75,76]

Additionally, the approach outlined in Eq. s (19)-(22) can be used to obtain random 3D orientations and compute an average mobility over more than $10^6$ random orientations. We plot this mobility $\mu_r$ in Figure 9c. Such a random mobility can be regarded as a reference mobility in a specimen with crystallites in several orientations, even though the carrier scattering at the boundaries between the crystallites is anyway neglected. The same analysis for the PF is shown in Figure 10.

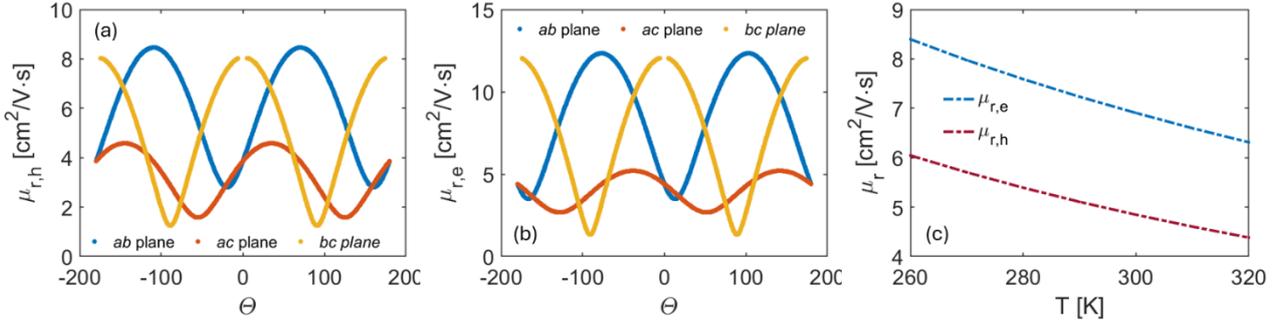

Figure 9: (a) anisotropic hole mobility in the crystal planes, as indicated; $\Theta$ is the angle, in deg, in respect of a cell axis. (b) anisotropic electron mobility in the crystal planes, as indicated. The average in-plane mobility is $\mu_{ab}$ ~ 6.1 (8.5) cm$^2$/Vs, $\mu_{ac}$ ~ 3.1 (4.3) cm$^2$/Vs, $\mu_{bc}$ ~ 5.7 (8.5) cm$^2$/Vs for holes (electrons). The in-plane anisotropy ratios are 3.0 (3.5), 2.9 (1.9), 6.5 (9.1), for the *ab*, *ac*, and *bc* planes, respectively, for holes (electrons). (c) average mobility computed for $2 \cdot 10^6$ random orientations of the electric field.

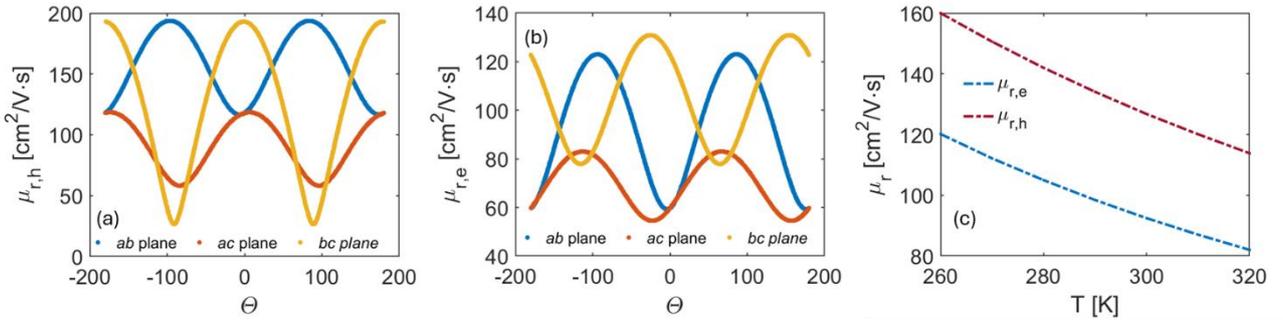

Figure 10: (a) anisotropic hole mobility in the crystal planes, as indicated; $\Theta$ is the angle, in degrees, with respect to the cell axis. (b) anisotropic electron mobility in the crystal planes, as indicated. The average in-plane mobility is $\mu_{ab}$ ~ 157.9 (94.3) cm$^2$/Vs, $\mu_{ac}$ ~ 91.2 (68.9) cm$^2$/Vs, $\mu_{bc}$ ~ 130.7 (104.4) cm$^2$/Vs for holes (electrons). The in-plane anisotropy ratios are 1.7 (2.1), 2.0 (1.5), 7.3 (1.7), for the *ab*, *ac*, and *bc* planes, respectively, for holes (electrons). (c) average mobility computed for $2 \cdot 10^6$ random orientations of the electric field.

We observe that the PF presents a much higher mobility. The idea that some polymorphs of acenes exhibit higher mobility is not new, to the extent that it has been suggested that the wide variability in mobility observed in pentacene OFETs can be attributed to the presence of different crystal forms, [77] and coexistence of structures has been detected. [9,78] So, the disorder is increased by the coexistence of polymorphs, with detrimental impact on the mobility. Thin films phases, also known as surface induced structures, [79] are particularly important in applications such as OFETs, as the OSC conducting channel lie at the interface with the dielectric layer. [80] In fact, pentacene thin-film structure has been reported to exhibit a higher mobility than the bulk. [81] The TF mobility here calculated is for a single crystal, i.e. for a 3D, highly pure system which, in fact, is actually experimentally unattainable, as this phase exists only within a few nm thickness before coexist with, or being replaced by, the bulk polymorph. [8] However, this result suggests that the thin-film phase exhibits improved charge transport characteristics.

Our goal now is to address the reasons behind the higher mobility of the PF. This can be due to the higher carrier band velocity, Figure 2, but mostly to the lower EPC strength, up to ~ 4× lower in the topmost VB and lowest CB – we recall that the scattering rate has a quadratic dependence on the EPC. We further challenge this task by addressing the average random mobility, the bandwidth, an average electronic structure quantity, the EPC, the interaction between the electronic structure and the vibrational modes. The bandwidth of the two CBs and the two VBs for each polymorph is reported in Figure 11a, where the $x$-axis is the average energy of the band. It is obvious that the PF has wider, more disperse, bands, which correspond to the larger velocity in Figure 2. The positive relationship between more disperse bands and mobility is demonstrated in Figure 11b, where the random mobility computed after Eq. s (19)-(22) is plotted in respect of the CB and VB bandwidth of the two polymorphs, here we considered the topmost VB and the lowest CB, denoted with '0' subscript, as they are the most involved in the charge transport. To gain additional insight into this qualitative consideration, we investigate the relationship between the random mobility at 300 K and the ratio between the bandwidth, which favors the mobility, and the effective EPC, for the band at the edge. We define an effective EPC strength with Eq. (23):

$$\text{EPC}_{B_0}^{eff} = \sqrt{\sum_\nu \frac{D^2_{n_0,n_0,\nu}}{\hbar\omega_\nu}} \quad (23)$$

where $D_{n,n,\nu}$ is the intra-band EPC for the band defined from Eq. (3) for the band $n_0$, which is the band at the edge (topmost VB or lowest CB), and the mode $\nu$, $\hbar\omega_\nu$ is the mode energy. This definition is based on the equation for the evaluation of the scattering rates, Eq. (11), and the effective EPC estimated with Eq. (23) has units of $\sqrt{eV}/\text{Å}$.

By plotting the random mobility at 300 K and the ratio between the bandwidth and such effective EPC, Figure 11c, a straightforward linear trend appears, represented by the fitting black dash-dot line with R-square = 0.9994. If we neglect the phonon frequency and sum only the EPC values, obtain a slightly worse trend with an R-square of 0.9938. Thus, the two key ingredients to have high mobility OSC: large bandwidth and weak effective EPC.

Thus, the core task becomes how to achieve weak EPC, and, in our case, what gives such lower EPC in PF. We believe that wider electronic bands, signatures of more stable interactions, can be a relevant hint. Also, the question of whether a larger BW signals a greater insensitivity of the transfer integral to the phonon modes, is worth deeper investigation.

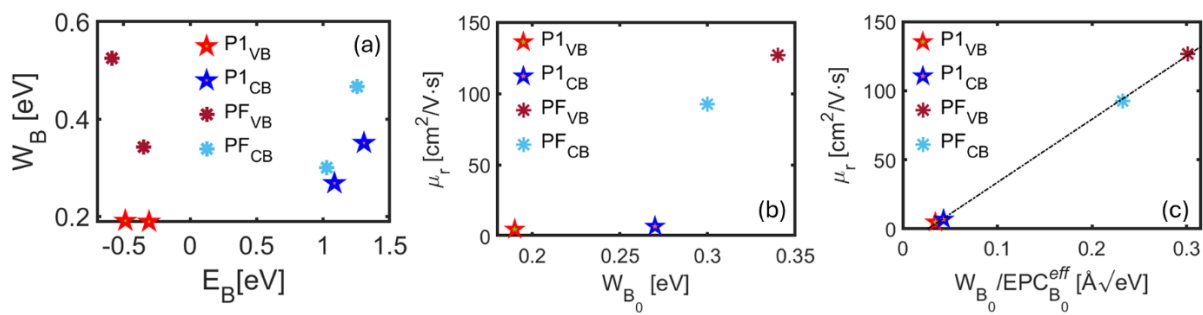

Figure 11: (a) total band width versus band energy, evaluated as the mean of the band eigenvalues, for the two valence bands (VB), bright red, and two conductions bands (CB), bright blue, of P1, and, with dark red and light blue, the same for PF. (b) dependence of the random mobility at 300 K, Eq. s (19)-(22), versus the average bandwidth of the two VBs or CBs for the two polymorphs. A clear monotonic dependence appears. (c) dependence on the random mobility at 300 K with respect to the ratio between the band width and the effective EPC, for the bands at the edge, i.e. the topmost VB and the lowest CB, which are the bands more involved in the transport. The effective EPC is defined as in Eq. (23). The dash-dot black line is the fitting line. A clear linear dependence appears, signaling the relevant parameters for mobility in OSC.

### d. Comparison with other methods

Finally, in order to benchmark the computational burden of our proposed method, for naphthalene we performed calculations using state-of-the-art DFPT methods. We followed the procedure described in the PERTURBO code, [31] which was already adopted in the literature to perform DFPT calculation of the mobility of naphthalene in literature, [82,83] labelled ph-wannier in Table III. We also used the EPA approximations, which has been never used for OSCs before. [32] The mobility computed with the ph-wannier method is already reported in literature, [82] and we don't replicate those data here. However, we highlight that in that case the mobility was overestimated by a factor of about 3, in line with our findings. The EPA method data are plotted in Figure 8c, red line. The computational settings used, as well as the core-hours employed, are reported in Table III. We observe the significantly reduced cost enabled by the proposed method, which means that it is possible to afford larger systems or to increase the HPC throughput.

| Method | $k$-grid | $q$-grid | 'tr2_ph' | VdW | core hours |
|---|---|---|---|---|---|
| this work | 3×5×3 | 2×3×2 | - | yes | 2,000 |
| EPA | 4×4×4 | 2×2×2 | $10^{-10}$ | no | 33,000 |
| ph-wannier | 4×4×4 | 2×4×2 | $10^{-12}$ | no | 19,500 |

Table III: comparison of the computational burden of the three methods discussed in the text. About the ph-wannier method, we followed the instructions in the PERTURBO manual up to the e-ph calculation and we used the grid as in ref. [83]. We didn't compute the mobility because the related mobility values are already reported in literature. The cutoff energy ('ecutwfc' parameter) is 70 Ry in Quantum Espresso ('ecutrho' parameter set to ten times the 'ecutwfc' parameter) and 900 eV in VASP. The 'tr2_ph' tag is the threshold for the self-consistent DFPT calculation, the threshold at the electronic structure level is always $10^{-8}$. In our proposed method, the $q$-grid must be regarded as the supercell size. The Van der Waals (VdW) interaction is always used at the electronic structure stage, not always at the phonon stage.

### IV. Conclusions

A method has been developed that combines DFT-level calculations and semiclassical transport simulations to compute the electron-phonon coupling in organic semiconductors. The model has been validated through a comparison of computed and experimental Raman spectra and single-crystal hole mobility for tetracene. The coupling has been estimated across the Brillouin zone for different phonon modes, and from these calculations, the charge mobility has been evaluated. The validity of the electronic structure and phonon calculations has been demonstrated by computing the low frequency Raman spectrum and comparing it with experimental results. Based on the achievement on tetracene, the suitability of the overall method has been confirmed through a comparison of computed and experimental single-crystal hole mobility of naphthalene and anthracene. A slightly overestimated mobility is obtained; however, including point ionized defects adjusts the mobility to fall within the experimental range, but at the cost of a worse agreement in its overall temperature dependence. This suggests that the development of a proper treatment for likely extended ionized defects is necessary to achieve accurate modeling of real specimens. Furthermore, the ratio between the bandwidth and the effective EPC is identified as a key parameter for ranking the mobility of OSCs.

We demonstrate that our approach is suitable for first-principles calculations of mobility in crystalline OSCs, at a significant lower computational burden in respect to DFPT approaches. Our work establishes a solid foundation for understanding how polymorphism impacts electronic properties and devises a simulation approach to link solid-state packing and charge transport.

**Acknowledgement**

We are thankful to prof. Alberto Girlando for fruitful discussions, to Dr. Andra Giunchi for supporting in the computational protocol for the Raman spectra. We acknowledge the CINECA award under the ISCRA initiative, for the availability of high-performance computing resources and support. We acknowledge funding from the European Union–Next-Generation EU via the Italian call PRIN 2022, project code 2022XZ2ZM8, "POLYPHON". T.S. thanks the Programma per Giovani Ricercatori "Rita Levi Montalcini" year 2020 (grant PGR20QN52R) of the Italian Ministry of University and Research (MUR) for the financial support. We acknowledge funding for the VASP license from the company M.M.B. s.r.l., Faenza (RA) Italy.

*Supplemental Material*

# Electron-phonon coupling and mobility modeling in organic semiconductors: method and application to tetracene polymorphs

Patrizio Graziosi[1,*], Raffaele Guido Della Valle[2], Tommaso Salzillo[2], Simone d'Agostino[3], Martina Zangari[2], Elisabetta Cané[2], Matteo Masino[4], Elisabetta Venuti[2,*]

[1] ISMN – CNR, Consiglio Nazionale delle Ricerche, via Gobetti 101, 40129, Bologna, Italy

[2] Dipartimento di Chimica Industriale "Toso Montanari", Università di Bologna, via Gobetti 85, 410129, Italy

[3] Dipartimento di Chimica "Giacomo Ciamician", Università di Bologna, Via F. Selmi, 2, 40126 Bologna, Italy

[4] Dipartimento di Scienze Chimiche, Della Vita e Della Sostenibilità Ambientale & INSTM-UdR Parma, Parco Area delle Scienze, 17/A, 43124 Parma, Italy

[*] patrizio.graziosi@cnr.it, elisabetta.venuti@unibo.it

# Contents

**A – Vibrational frequencies in the lattice modes range for P1 with the symmetry assignment, and comparison with the experimental values from ref. [53] of the main text**

Table S1: Vibrational Frequencies and symmetry assignment for P1 lattice phonon. Values in red are above the lattice mode range and they are reported only for comparison with experiments.

| Symmetry | Freq. calc. [cm$^{-1}$] | Freq. exp. [cm$^{-1}$] |
|---|---|---|
| Au | 33.2 | |
| Ag | 45.3 | 42.3 |
| Ag | 48.7 | 47.8 |
| Ag | 60.5 | 58.5 |
| Au | 66.5 | |
| Au | 69.4 | |
| Ag | 86.2 | 88.4 |
| Au | 98.4 | |
| Au | 103.4 | |
| Ag | 117.4 | 117.1 |
| Ag | 128.6 | 129.8 |
| Au | 136.1 | |
| Au | 144.5 | |
| Ag | 164.4 | |
| Au | 164.6 | |
| Ag | 168.2 | 168.2 |
| Au | 169.3 | |
| Ag | 208.3 | 211.2 |
| Ag | 213.5 | 217.0 |

**B – EPC values**

Tables with modes frequency in the different points of the BZ and EPC values are reported. The average atom displacements are around 5.9×10$^{-4}$ Å for the inter-molecular mode and around 5.8×10$^{-4}$ Å for the intra-molecular modes. The order of the bands is referred to the edge at the gap; the 1$^{st}$ VB is the one with the highest maximum whereas the 1$^{st}$ CB is the one with the lowest minimum.

Thus, for the tetracene polymorph P1, about the inter-molecular modes, which are the lowest 16 modes mostly involved in the charge transport limitations, the frequency values are reported in Table S2, the EPC values for the VB are listed in Tables S3 (intra-band in the topmost VB) S4 (intra-band in the lower VB), and S4 (inter-band), and for the CB in Tables S6 (intra-band in the lowest CB), S7 (intra-band in the upper CB), and S8 (inter-band). We also report the EPC for the intra-molecular modes up to 400 cm$^{-1}$. The frequency are listed in Table S9, the EPC values for the VB are reported in Tables S9, S11 and S12 and for the CB in S12, S13, and S15, in the same way of the inter-molecular: two tables for the intra-band (one for each band) and a table for the inter-band EPC.

In a similar way, for the thin film polymorph, we report in Table S16 the mode and *q*-point specific frequencies for the inter-molecular modes, in Tables S17, S18, and S19 the EPC values for the VB for the intra-band processes of the topmost VB, the lower VB, and the inter-band processes, respectively, and in Tables S20, S21, and S22 the EPC values for the CB for the intra-band processes of the lowest CB, the other CB, and the inter-band processes, respectively. Then, we report the same information for the intra-molecular modes up to 400 cm$^{-1}$: the frequency values are listed in Table S23, the EPC values for intra-band and inter-band processes for the VB are reported in Tables S24, S25, and S26, while Table S27 and S28 contain the EPC for the intra-band processes of the two CB and Table 28 reports the inter-band EPC values for the CB.

In Table S29, the mode and *q*-point pairs with the highest EPC values are highlighted in green. These are the Brillouin Zone points and modes which have the most impact on the hole transport in the P1 polymorph.

We report the EPC values DOS-weighted averaged across the Brillouin Zone (BZ) in Tables S30 and S31 for P1 and PF polymorphs, respectively.

The codes used to evaluate the EPC can be found at https://github.com/PatrizioGraziosi/EPHOS.

***Tetracene – P1***

Inter-molecular modes

Table S2: frequencies in cm$^{-1}$

| Γ | Z – *c* | Y - *b* | X - *a* | V | U | T | R |
|---|---|---|---|---|---|---|---|
| -0.25 | 18.18 | 28.89 | 33.96 | 24.76 | 26.1 | 22.91 | 31.59 |
| -0.16 | 21.79 | 34.2 | 42.45 | 28.44 | 29.36 | 22.99 | 38.66 |
| 0.17 | 23.36 | 36.62 | 44.57 | 48.19 | 46.56 | 37.07 | 40.96 |
| 33.18 | 24.43 | 41 | 51.78 | 49.89 | 52.9 | 47.75 | 45.36 |
| 45.35 | 31.61 | 48.12 | 56.74 | 50.01 | 56.22 | 47.78 | 53.85 |
| 48.74 | 41.47 | 53.45 | 60.3 | 54.75 | 59.3 | 51.55 | 57.52 |
| 60.46 | 59.37 | 56.73 | 61.69 | 75.48 | 64.11 | 62.26 | 70.96 |
| 66.47 | 63.77 | 59.49 | 69.11 | 76.71 | 66.78 | 62.94 | 80.31 |
| 69.35 | 89.83 | 73.88 | 72.53 | 85.13 | 78.88 | 66.73 | 84.75 |
| 86.17 | 92.82 | 76.47 | 79.04 | 87.34 | 81.29 | 67.02 | 86.62 |
| 98.43 | 95.91 | 93.93 | 92.07 | 87.4 | 91.19 | 92.63 | 91.01 |
| 103.39 | 105.75 | 97.19 | 97 | 90.57 | 93.76 | 96.89 | 92.91 |
| 117.44 | 119.69 | 121.77 | 104.91 | 93.53 | 102.46 | 121.91 | 96.81 |
| 128.65 | 129.1 | 131.8 | 106.59 | 99.62 | 103.31 | 130.04 | 103.95 |
| 136.14 | 134.17 | 136.34 | 125.98 | 120.93 | 122.99 | 138.18 | 122.46 |
| 144.48 | 143.38 | 141.79 | 129.18 | 133.9 | 128.63 | 144.04 | 128.23 |

Table S2: 1$^{st}$ VB, EPC in eV/Å, the largest couplings are highlighted in green, darker the stronger.

| Γ | Z – *c* | Y - *b* | X - *a* | V | U | T | R |
|---|---|---|---|---|---|---|---|
| 0.02 | 0.01 | 0.02 | 0.16 | 0.12 | 0.01 | 0.03 | 0.02 |
| 0.01 | 0.01 | 0.25 | 0.02 | 0.55 | 0.01 | 0.02 | 0.09 |
| 0.02 | 0.1 | 0.02 | 0.35 | 0.18 | 0.19 | 0.01 | 0.01 |
| 0.02 | 0.01 | 0.04 | 0 | 0.14 | 0.01 | 0.01 | 0.14 |
| 0.01 | 0.17 | 0.01 | 0.01 | 0.01 | 0.12 | 0.04 | 0.01 |
| 0.03 | 0.13 | 0.07 | 0.01 | 0.01 | 0.01 | 0.22 | 0.01 |
| 0.1 | 0.01 | 0.01 | 0.1 | 0.02 | 0.01 | 0.12 | 0.07 |
| 0.01 | 0.06 | 0.15 | 0.01 | 0.06 | 0.19 | 0.01 | 0.06 |
| 0.01 | 0.1 | 0.01 | 0.01 | 0.01 | 0.01 | 0.13 | 0.02 |
| 0.02 | 0.02 | 0.12 | 0.06 | 0.12 | 0.01 | 0.02 | 0.05 |
| 0.01 | 0.05 | 0.05 | 0.01 | 0.01 | 0.02 | 0.05 | 0.06 |
| 0.02 | 0.07 | 0.02 | 0.03 | 0.01 | 0.06 | 0.01 | 0.02 |
| 0.03 | 0.01 | 0.01 | 0.09 | 0.03 | 0.02 | 0.05 | 0.02 |
| 0.01 | 0.01 | 0.04 | 0.01 | 0.1 | 0.02 | 0.01 | 0.01 |
| 0.02 | 0.03 | 0.02 | 0.02 | 0.03 | 0.08 | 0.01 | 0.01 |
| 0.01 | 0.05 | 0.01 | 0.04 | 0.06 | 0.01 | 0.04 | 0.02 |

Table S3: 2nd VB, EPC in eV/Å

| Γ | Z – *c* | Y - *b* | X - *a* | V | U | T | R |
|---|---|---|---|---|---|---|---|
| 0.02 | 0.02 | 0.02 | 0.03 | 0.03 | 0.01 | 0.26 | 0.03 |
| 0.02 | 0.26 | 0.06 | 0.02 | 0.33 | 0.29 | 0.03 | 0.34 |
| 0.02 | 0.16 | 0.03 | 0.44 | 0.3 | 0.18 | 0.01 | 0.02 |
| 0.03 | 0.02 | 0.1 | 0.01 | 0.15 | 0.02 | 0.01 | 0.18 |
| 0.06 | 0.11 | 0.01 | 0.06 | 0.02 | 0.34 | 0.12 | 0.01 |
| 0.07 | 0.13 | 0.06 | 0.02 | 0.01 | 0.01 | 0.04 | 0.02 |
| 0.12 | 0.01 | 0.01 | 0.21 | 0.03 | 0.02 | 0.03 | 0.07 |
| 0.01 | 0.17 | 0.07 | 0.02 | 0.04 | 0.36 | 0.01 | 0.12 |
| 0.02 | 0.08 | 0.02 | 0.02 | 0.01 | 0.06 | 0.01 | 0.02 |
| 0 | 0.02 | 0.07 | 0.1 | 0.15 | 0.01 | 0.02 | 0.05 |
| 0.02 | 0.12 | 0.04 | 0.02 | 0.01 | 0.02 | 0.04 | 0.11 |
| 0.02 | 0.06 | 0.03 | 0.02 | 0.01 | 0.03 | 0.02 | 0.02 |
| 0.09 | 0.01 | 0.02 | 0.01 | 0.22 | 0.02 | 0.08 | 0.02 |
| 0.22 | 0.02 | 0.11 | 0.01 | 0.06 | 0.18 | 0.02 | 0.01 |
| 0.02 | 0.1 | 0.14 | 0.02 | 0.06 | 0.04 | 0.01 | 0.01 |
| 0.02 | 0.19 | 0.02 | 0.18 | 0.09 | 0.01 | 0.2 | 0.03 |

Table S4: Davydov splitting, VB, EPC in eV/Å

| Γ | Z – *c* | Y - *b* | X - *a* | V | U | T | R |
|---|---|---|---|---|---|---|---|
| 0 | 0.01 | 0 | 0.21 | 0.17 | 0.01 | 0.25 | 0.01 |
| 0 | 0.19 | 0.03 | 0 | 0.21 | 0.01 | 0.01 | 0.04 |
| 0.01 | 0.07 | 0.01 | 0.02 | 0.05 | 0.04 | 0 | 0.01 |
| 0.01 | 0 | 0.22 | 0 | 0.22 | 0 | 0 | 0.09 |
| 0.04 | 0.14 | 0 | 0.08 | 0.01 | 0.03 | 0.24 | 0 |
| 0.08 | 0.15 | 0.12 | 0 | 0 | 0 | 0.03 | 0.01 |
| 0.01 | 0 | 0 | 0.08 | 0.01 | 0 | 0.02 | 0.16 |
| 0 | 0.14 | 0.29 | 0.01 | 0.04 | 0.15 | 0 | 0.15 |
| 0.01 | 0 | 0 | 0 | 0 | 0.24 | 0.11 | 0.01 |
| 0.04 | 0 | 0.45 | 0.14 | 0.15 | 0 | 0.01 | 0.41 |
| 0 | 0.19 | 0.56 | 0 | 0 | 0 | 0.23 | 0.3 |
| 0 | 0.11 | 0.01 | 0.25 | 0 | 0.24 | 0.01 | 0 |
| 0.32 | 0 | 0 | 0.43 | 0.12 | 0.01 | 0.23 | 0 |
| 0.91 | 0.01 | 0.44 | 0 | 0.87 | 0.33 | 0 | 0 |
| 0 | 0.24 | 0.44 | 0.01 | 0.25 | 0.12 | 0 | 0 |
| 0 | 0.77 | 0 | 0.65 | 0.27 | 0 | 0.74 | 0.01 |

Table S6: 1st CB, EPC in eV/Å

| Γ | Z – *c* | Y - *b* | X - *a* | V | U | T | R |
|---|---|---|---|---|---|---|---|
| 0.03 | 0.03 | 0.03 | 0.17 | 0.28 | 0.03 | 0.1 | 0.03 |
| 0.02 | 0.05 | 0.19 | 0.03 | 0.34 | 0.25 | 0.04 | 0.38 |
| 0.03 | 0 | 0.04 | 0.44 | 0.33 | 0.15 | 0.02 | 0.03 |
| 0.04 | 0.03 | 0.02 | 0.02 | 0.14 | 0.03 | 0.03 | 0.07 |
| 0.03 | 0.17 | 0.02 | 0.06 | 0.03 | 0.28 | 0.01 | 0.03 |
| 0.06 | 0.11 | 0.07 | 0.03 | 0.03 | 0.03 | 0.16 | 0.03 |
| 0.16 | 0.02 | 0.02 | 0.07 | 0.04 | 0.03 | 0 | 0.06 |
| 0.02 | 0.08 | 0.07 | 0.03 | 0.06 | 0.22 | 0.03 | 0.11 |
| 0.03 | 0.07 | 0.03 | 0.02 | 0.03 | 0.04 | 0.04 | 0.04 |
| 0.04 | 0.03 | 0.08 | 0.04 | 0.09 | 0.03 | 0.03 | 0.1 |
| 0.03 | 0.14 | 0.05 | 0.03 | 0.03 | 0.03 | 0.1 | 0.06 |
| 0.04 | 0 | 0.04 | 0.06 | 0.02 | 0.04 | 0.03 | 0.03 |
| 0 | 0.03 | 0.03 | 0.07 | 0.17 | 0.03 | 0.05 | 0.03 |
| 0.15 | 0.02 | 0.13 | 0.03 | 0.04 | 0.03 | 0.03 | 0.02 |
| 0.03 | 0.05 | 0.08 | 0.03 | 0.06 | 0.1 | 0.02 | 0.02 |
| 0.03 | 0.07 | 0.03 | 0.05 | 0.15 | 0.03 | 0.1 | 0.03 |

Table S7: 2nd CB, EPC in eV/Å

| Γ | Z – *c* | Y - *b* | X - *a* | V | U | T | R |
|---|---|---|---|---|---|---|---|
| 0.04 | 0.03 | 0.04 | 0.36 | 0.46 | 0.03 | 0.25 | 0.04 |
| 0.03 | 0.05 | 0.5 | 0.04 | 0.97 | 0.15 | 0.05 | 0.36 |
| 0.04 | 0.02 | 0.05 | 0.72 | 0.44 | 0.16 | 0.03 | 0.03 |
| 0.04 | 0.03 | 0.17 | 0.02 | 0.27 | 0.03 | 0.04 | 0.02 |
| 0.19 | 0.2 | 0.03 | 0.07 | 0.03 | 0.32 | 0 | 0.03 |
| 0 | 0.16 | 0.13 | 0.03 | 0.03 | 0.03 | 0.19 | 0.03 |
| 0.05 | 0.03 | 0.03 | 0.14 | 0.04 | 0.03 | 0.16 | 0.08 |
| 0.02 | 0.07 | 0.36 | 0.03 | 0.06 | 0.3 | 0.03 | 0.08 |
| 0.03 | 0.17 | 0.04 | 0.03 | 0.03 | 0.05 | 0.15 | 0.04 |
| 0.02 | 0.04 | 0.01 | 0.14 | 0.2 | 0.03 | 0.03 | 0.09 |
| 0.03 | 0.15 | 0.21 | 0.03 | 0.03 | 0.04 | 0.07 | 0.04 |
| 0.04 | 0.03 | 0.04 | 0.11 | 0.03 | 0.12 | 0.03 | 0.03 |
| 0.16 | 0.04 | 0.03 | 0.01 | 0.14 | 0.04 | 0.11 | 0.04 |
| 0.07 | 0.03 | 0.03 | 0.03 | 0.02 | 0.15 | 0.03 | 0.03 |
| 0.04 | 0.18 | 0.09 | 0.04 | 0.03 | 0.09 | 0.03 | 0.03 |
| 0.03 | 0.06 | 0.03 | 0.03 | 0.08 | 0.03 | 0.09 | 0.04 |

Table S8: Davydov splitting, CB, EPC in eV/Å

| Γ | Z – *c* | Y - *b* | X - *a* | V | U | T | R |
|---|---|---|---|---|---|---|---|
| 0.07 | 0.06 | 0.07 | 0.21 | 0.42 | 0.06 | 0.05 | 0.08 |
| 0.05 | 0.12 | 0.52 | 0.08 | 1.07 | 0.62 | 0.08 | 0.78 |
| 0.07 | 0.07 | 0.09 | 1.17 | 0.7 | 0.54 | 0.05 | 0.06 |
| 0.08 | 0.06 | 0.33 | 0.05 | 0.14 | 0.07 | 0.07 | 0.39 |
| 0.05 | 0.25 | 0.06 | 0.08 | 0.07 | 0.59 | 0.07 | 0.06 |
| 0.06 | 0.21 | 0.08 | 0.07 | 0.06 | 0.06 | 0.29 | 0.06 |
| 0.15 | 0.05 | 0.05 | 0.22 | 0.08 | 0.07 | 0.01 | 0.2 |
| 0.05 | 0.07 | 0.18 | 0.07 | 0.24 | 0.51 | 0.06 | 0.28 |
| 0.06 | 0 | 0.07 | 0.06 | 0.06 | 0 | 0.17 | 0.08 |
| 0.13 | 0.07 | 0.14 | 0 | 0.28 | 0.06 | 0.07 | 0.24 |
| 0.05 | 0.16 | 0.14 | 0.06 | 0.06 | 0.07 | 0.05 | 0.07 |
| 0.07 | 0.08 | 0.09 | 0.02 | 0.05 | 0.08 | 0.06 | 0.07 |
| 0.01 | 0.07 | 0.07 | 0.18 | 0.44 | 0.08 | 0.13 | 0.07 |
| 0.34 | 0.06 | 0.34 | 0.06 | 0.21 | 0.07 | 0.06 | 0.05 |
| 0.07 | 0.03 | 0.17 | 0.07 | 0.1 | 0.18 | 0.05 | 0.05 |
| 0.06 | 0.13 | 0.06 | 0.07 | 0.31 | 0.06 | 0.24 | 0.07 |

***Tetracene – P1***

Intra-molecular modes up to 400 cm$^{-1}$

Table S9: Frequencies in cm$^{-1}$

| Γ | Z – ***c*** | Y - ***b*** | X - ***a*** | V | U | T | R |
|---|---|---|---|---|---|---|---|
| 164.39 | 152.68 | 160.8 | 159.35 | 164.2 | 164.15 | 163.49 | 156.88 |
| 164.55 | 166.88 | 161.09 | 159.87 | 164.73 | 164.36 | 164.77 | 162.9 |
| 168.21 | 167.47 | 169.11 | 167.12 | 167.17 | 168.12 | 168.37 | 165.91 |
| 169.27 | 168.92 | 169.53 | 169.88 | 167.22 | 168.46 | 168.55 | 169.21 |
| 208.34 | 210.91 | 211.11 | 206.85 | 206.42 | 207.31 | 210.63 | 206.82 |
| 213.55 | 213.7 | 213 | 212.77 | 210.82 | 210.45 | 210.83 | 214.83 |
| 268.05 | 271.92 | 272.28 | 271.96 | 268.23 | 269.7 | 270.84 | 271.74 |
| 274.27 | 275.65 | 275.29 | 273.19 | 271.34 | 270.15 | 272.38 | 273.13 |
| 298.9 | 297.72 | 298.18 | 297.67 | 298.11 | 298.53 | 299.01 | 297.67 |
| 299.71 | 299.96 | 299.47 | 299.02 | 299.59 | 299.19 | 299.58 | 299 |
| 314.73 | 313.32 | 313.7 | 313.57 | 315.26 | 314.99 | 315.19 | 313.77 |
| 318.36 | 315.42 | 314.51 | 314.78 | 318.28 | 318.28 | 317.92 | 314.88 |
| 320.57 | 317.95 | 318.93 | 322.96 | 322.91 | 323.25 | 321.3 | 322.16 |
| 322.47 | 322.32 | 321.04 | 323.89 | 327.21 | 326.4 | 321.37 | 324.62 |
| 381.09 | 380.64 | 380.72 | 375.05 | 375.26 | 375.94 | 381.22 | 373.99 |
| 385.28 | 384.59 | 384.62 | 376.37 | 377.12 | 376.7 | 385.34 | 377.12 |

Table S10: 1$^{st}$ VB, EPC in eV/Å

| Γ | Z – ***c*** | Y - ***b*** | X - ***a*** | V | U | T | R |
|---|---|---|---|---|---|---|---|
| 0.06 | 0.01 | 0.01 | 0.01 | 0.01 | 0.14 | 0.01 | 0.01 |
| 0 | 0.01 | 0.01 | 0.01 | 0.12 | 0 | 0.07 | 0.03 |
| 0.02 | 0.09 | 0.01 | 0.05 | 0.01 | 0.09 | 0 | 0.01 |
| 0 | 0.05 | 0.04 | 0 | 0.06 | 0 | 0.13 | 0.01 |
| 0.02 | 0 | 0 | 0.05 | 0 | 0.01 | 0.01 | 0.02 |
| 0.01 | 0 | 0.06 | 0 | 0.01 | 0.1 | 0.02 | 0.1 |
| 0.01 | 0.07 | 0.07 | 0.01 | 0.02 | 0.01 | 0.01 | 0 |
| 0.01 | 0.02 | 0.01 | 0.03 | 0.06 | 0.07 | 0.02 | 0.01 |
| 0.01 | 0 | 0 | 0.05 | 0 | 0.2 | 0.01 | 0.09 |
| 0.04 | 0 | 0.02 | 0 | 0.01 | 0 | 0.18 | 0.18 |
| 0.11 | 0.01 | 0.1 | 0.01 | 0 | 0.16 | 0.01 | 0.1 |
| 0.04 | 0 | 0 | 0.27 | 0.01 | 0.02 | 0.24 | 0.59 |
| 0 | 0.01 | 0 | 0 | 0 | 0.03 | 0.05 | 0.01 |
| 0 | 0.01 | 0.01 | 0.1 | 0.09 | 0 | 0 | 0 |
| 0.05 | 0 | 0 | 0.01 | 0.01 | 0.02 | 0.01 | 0 |
| 0 | 0 | 0 | 0.09 | 0.01 | 0 | 0.01 | 0.07 |

Table S11: 2nd VB, EPC in eV/Å

| Γ | Z – *c* | Y - *b* | X - *a* | V | U | T | R |
|---|---|---|---|---|---|---|---|
| 0.01 | 0.01 | 0.01 | 0.01 | 0.01 | 0.01 | 0.01 | 0.08 |
| 0 | 0.01 | 0.01 | 0.03 | 0.21 | 0 | 0.02 | 0.07 |
| 0 | 0.14 | 0.01 | 0.07 | 0.01 | 0.01 | 0 | 0.01 |
| 0 | 0.15 | 0.05 | 0 | 0.06 | 0 | 0.09 | 0.01 |
| 0.08 | 0 | 0 | 0.11 | 0 | 0.01 | 0.01 | 0.05 |
| 0.04 | 0 | 0.09 | 0.01 | 0.01 | 0.18 | 0.1 | 0.12 |
| 0.01 | 0 | 0.04 | 0.02 | 0.06 | 0.01 | 0.01 | 0 |
| 0.01 | 0.01 | 0.01 | 0.03 | 0.03 | 0.11 | 0.03 | 0.01 |
| 0 | 0 | 0.01 | 0.08 | 0 | 0.07 | 0.01 | 0.15 |
| 0.09 | 0 | 0.05 | 0.01 | 0.01 | 0 | 0.13 | 0.08 |
| 0.15 | 0.01 | 0.1 | 0.01 | 0 | 0.01 | 0.01 | 0.03 |
| 0.08 | 0 | 0 | 0.2 | 0.01 | 0.03 | 0.06 | 0.27 |
| 0.01 | 0.08 | 0 | 0 | 0.03 | 0.12 | 0.07 | 0.01 |
| 0 | 0.02 | 0.06 | 0.19 | 0.15 | 0 | 0.01 | 0 |
| 0.02 | 0 | 0.01 | 0.01 | 0.01 | 0.12 | 0.2 | 0.05 |
| 0.01 | 0.01 | 0.02 | 0.11 | 0.01 | 0 | 0.01 | 0.07 |

Table S12: Davydov splitting, VB, EPC in eV/Å

| Γ | Z – *c* | Y - *b* | X - *a* | V | U | T | R |
|---|---|---|---|---|---|---|---|
| 0.02 | 0 | 0.01 | 0 | 0 | 0.09 | 0.01 | 0.15 |
| 0 | 0 | 0 | 0.02 | 0.09 | 0 | 0.32 | 0.58 |
| 0.03 | 0.02 | 0.01 | 0.04 | 0 | 0.33 | 0 | 0 |
| 0 | 0.01 | 0.02 | 0 | 0.02 | 0 | 0.01 | 0 |
| 0.11 | 0 | 0 | 0.68 | 0 | 0 | 0 | 0.08 |
| 0.33 | 0 | 0.17 | 0 | 0 | 0.37 | 0.28 | 0.2 |
| 0 | 0.04 | 0.3 | 0.01 | 0.11 | 0 | 0.01 | 0 |
| 0 | 0.01 | 0 | 0.24 | 0.14 | 0.24 | 0.04 | 0 |
| 0.06 | 0 | 0 | 0.04 | 0 | 0.16 | 0.01 | 0.11 |
| 0.04 | 0 | 0.08 | 0 | 0 | 0 | 0.06 | 0.18 |
| 0.01 | 0 | 0 | 0 | 0 | 0.09 | 0 | 0.25 |
| 0.02 | 0 | 0 | 0 | 0 | 0.01 | 0.05 | 0.04 |
| 0 | 0.03 | 0 | 0 | 0.02 | 0.8 | 0.47 | 0 |
| 0 | 0.66 | 0.16 | 0.17 | 0.15 | 0 | 0.01 | 0 |
| 0.02 | 0 | 0 | 0 | 0 | 0.54 | 0.79 | 0.16 |
| 0.01 | 0 | 0.03 | 0.22 | 0 | 0 | 0 | 0.56 |

Table S13: 1st CB, EPC in eV/Å

| Γ | Z – *c* | Y - *b* | X - *a* | V | U | T | R |
|---|---|---|---|---|---|---|---|
| 0.08 | 0.01 | 0.08 | 0.01 | 0.02 | 0.27 | 0.01 | 0.05 |
| 0.01 | 0.01 | 0.01 | 0 | 0.32 | 0.01 | 0.1 | 0.05 |
| 0.01 | 0.11 | 0.02 | 0 | 0.01 | 0.19 | 0.01 | 0.01 |
| 0.01 | 0.04 | 0 | 0.01 | 0.03 | 0.01 | 0.16 | 0.01 |
| 0.03 | 0.01 | 0.01 | 0.05 | 0.01 | 0.01 | 0.01 | 0.05 |
| 0.03 | 0.01 | 0.03 | 0 | 0.01 | 0.05 | 0.06 | 0.06 |
| 0.01 | 0.02 | 0.05 | 0.02 | 0.01 | 0.01 | 0.02 | 0.01 |
| 0.01 | 0.01 | 0.01 | 0.07 | 0.03 | 0.02 | 0.05 | 0.01 |
| 0.01 | 0.01 | 0.01 | 0.12 | 0.01 | 0 | 0.01 | 0.1 |
| 0.01 | 0.01 | 0.01 | 0.01 | 0.01 | 0.01 | 0.05 | 0.03 |
| 0.41 | 0.01 | 0.33 | 0.02 | 0.01 | 0.43 | 0.01 | 0.36 |
| 0.16 | 0 | 0.01 | 0.27 | 0.01 | 0.02 | 0.36 | 0.35 |
| 0.01 | 0.04 | 0.01 | 0.01 | 0.05 | 0.01 | 0.05 | 0.01 |
| 0.01 | 0.01 | 0.07 | 0.09 | 0.08 | 0.01 | 0.01 | 0.01 |
| 0.01 | 0.01 | 0.01 | 0.01 | 0.01 | 0.09 | 0.05 | 0.05 |
| 0.04 | 0 | 0.05 | 0.06 | 0.01 | 0.01 | 0.01 | 0.02 |

Table S14: 2nd CB, EPC in eV/Å

| Γ | Z – *c* | Y - *b* | X - *a* | V | U | T | R |
|---|---|---|---|---|---|---|---|
| 0.05 | 0.01 | 0.03 | 0.02 | 0.02 | 0.47 | 0.02 | 0.01 |
| 0.01 | 0.01 | 0.01 | 0.12 | 0.34 | 0.01 | 0 | 0.04 |
| 0.01 | 0.09 | 0.02 | 0.01 | 0.01 | 0.13 | 0.01 | 0.01 |
| 0.01 | 0.22 | 0.07 | 0 | 0.1 | 0.01 | 0 | 0.01 |
| 0.05 | 0.01 | 0 | 0.08 | 0.01 | 0.02 | 0.02 | 0 |
| 0.07 | 0.01 | 0.06 | 0 | 0.02 | 0.04 | 0.15 | 0.02 |
| 0.02 | 0.01 | 0.01 | 0.02 | 0.05 | 0.02 | 0.02 | 0.01 |
| 0.01 | 0.07 | 0.02 | 0.1 | 0 | 0.02 | 0.07 | 0.01 |
| 0 | 0.01 | 0.01 | 0.11 | 0.01 | 0.11 | 0.01 | 0.01 |
| 0.02 | 0.01 | 0.02 | 0.01 | 0.01 | 0.01 | 0.09 | 0.06 |
| 0.45 | 0.02 | 0.39 | 0.02 | 0.01 | 0.64 | 0.02 | 0.6 |
| 0.19 | 0 | 0.01 | 0.48 | 0.01 | 0.03 | 0.86 | 0.98 |
| 0.01 | 0.1 | 0.01 | 0.01 | 0 | 0.21 | 0.14 | 0.02 |
| 0 | 0.06 | 0.15 | 0.12 | 0.07 | 0 | 0.01 | 0 |
| 0.05 | 0.01 | 0.01 | 0.01 | 0.01 | 0.11 | 0.1 | 0.12 |
| 0.09 | 0 | 0.09 | 0.06 | 0.02 | 0.01 | 0.02 | 0.01 |

Table S15: Davydov splitting, CB, EPC in eV/Å

| Γ | Z – *c* | Y - *b* | X - *a* | V | U | T | R |
|---|---|---|---|---|---|---|---|
| 0.05 | 0.03 | 0.09 | 0.03 | 0.04 | 0.7 | 0.03 | 0.07 |
| 0.02 | 0.02 | 0.03 | 0.12 | 0.75 | 0.02 | 0.01 | 0.2 |
| 0.07 | 0.47 | 0.04 | 0.06 | 0.03 | 0.44 | 0.01 | 0.02 |
| 0.01 | 0.05 | 0.07 | 0.01 | 0.09 | 0.01 | 0.58 | 0.02 |
| 0.01 | 0.02 | 0 | 0.01 | 0.02 | 0.03 | 0.03 | 0.01 |
| 0.01 | 0.01 | 0.11 | 0.01 | 0.03 | 0.1 | 0.01 | 0.09 |
| 0.03 | 0.09 | 0.09 | 0.05 | 0.05 | 0.03 | 0.04 | 0.03 |
| 0.03 | 0.07 | 0.03 | 0.12 | 0.11 | 0.15 | 0.09 | 0.02 |
| 0.02 | 0.02 | 0.02 | 0.09 | 0.01 | 0.01 | 0.03 | 0.14 |
| 0.02 | 0.01 | 0.03 | 0.02 | 0.03 | 0.02 | 0.02 | 0.06 |
| 0.85 | 0.03 | 0.81 | 0.04 | 0.02 | 0.74 | 0.03 | 0.72 |
| 0.35 | 0 | 0.02 | 0.73 | 0.03 | 0.06 | 1.08 | 1.13 |
| 0.02 | 0.07 | 0.02 | 0.01 | 0.06 | 0.14 | 0.17 | 0.03 |
| 0.01 | 0.1 | 0.06 | 0.19 | 0.09 | 0.01 | 0.02 | 0.01 |
| 0.09 | 0.02 | 0.03 | 0.02 | 0.02 | 0.25 | 0.04 | 0.18 |
| 0.06 | 0 | 0.07 | 0.13 | 0.03 | 0.01 | 0.03 | 0.01 |

***Tetracene – film phase***

Inter-molecular modes

Table S16: Frequencies in cm$^{-1}$

| Γ | Z – *c* | Y - *b* | X - *a* | V | U | T | R |
|---|---|---|---|---|---|---|---|
| -0.27 | 7.46 | 23.45 | 21.41 | 16.23 | 31.75 | 31.1 | 34.28 |
| -0.19 | 12.86 | 25.73 | 22.94 | 25.83 | 34.15 | 33.93 | 35.99 |
| -0.12 | 18.97 | 35.56 | 43.07 | 44.5 | 44.85 | 34.86 | 38.33 |
| 16.28 | 24.15 | 42.71 | 49.1 | 45.49 | 52.37 | 37.82 | 40.74 |
| 37.14 | 36.7 | 47.95 | 52.58 | 45.91 | 53.87 | 47.09 | 48.8 |
| 51.58 | 40.36 | 48.75 | 56.98 | 48.81 | 55.09 | 48.46 | 51.62 |
| 63.49 | 52.91 | 60.48 | 63.85 | 79.75 | 67.13 | 60.58 | 78.41 |
| 64.84 | 67.3 | 60.79 | 68.12 | 84.64 | 68.53 | 61.95 | 82.33 |
| 68.14 | 92.87 | 73.19 | 86.03 | 85 | 74.29 | 69.69 | 84.46 |
| 93.06 | 94.8 | 74.57 | 86.97 | 89.98 | 78.52 | 70.65 | 87.78 |
| 94.49 | 99.61 | 84.69 | 92.54 | 91.36 | 96.8 | 101.98 | 89.73 |
| 100.77 | 105.41 | 91.7 | 96.53 | 92.4 | 99.33 | 103.18 | 90.14 |
| 116.49 | 121.05 | 123.03 | 96.58 | 92.5 | 107.43 | 124.92 | 104.58 |
| 124.5 | 124.16 | 123.56 | 100.19 | 93.81 | 110.53 | 125.93 | 109.6 |
| 146.52 | 139.31 | 147.62 | 122.91 | 121.44 | 120.66 | 143 | 117.6 |
| 149.01 | 147.36 | 149.12 | 123.71 | 126 | 124.52 | 144.47 | 124.48 |

Table S17: 1$^{st}$ VB, EPC in eV/Å

| Γ | Z – *c* | Y - *b* | X - *a* | V | U | T | R |
|---|---|---|---|---|---|---|---|
| 0 | 0 | 0 | 0 | 0 | 0 | 0 | 0.06 |
| 0 | 0.04 | 0.04 | 0.01 | 0 | 0.04 | 0.02 | 0 |
| 0 | 0 | 0 | 0 | 0 | 0 | 0 | 0.04 |
| 0.03 | 0.01 | 0.02 | 0.02 | 0.01 | 0.02 | 0.06 | 0 |
| 0 | 0 | 0 | 0 | 0 | 0 | 0 | 0.03 |
| 0 | 0.02 | 0.09 | 0.08 | 0.01 | 0.04 | 0.04 | 0.01 |
| 0 | 0 | 0 | 0 | 0.02 | 0.05 | 0 | 0 |
| 0 | 0 | 0.03 | 0.01 | 0.01 | 0 | 0.01 | 0 |
| 0 | 0 | 0 | 0.05 | 0 | 0.05 | 0 | 0 |
| 0 | 0 | 0 | 0 | 0 | 0 | 0.01 | 0.04 |
| 0 | 0 | 0.01 | 0.02 | 0 | 0 | 0 | 0 |
| 0 | 0 | 0 | 0.04 | 0 | 0.01 | 0 | 0.03 |
| 0.01 | 0 | 0 | 0 | 0.03 | 0.01 | 0 | 0.03 |
| 0.01 | 0 | 0.01 | 0 | 0.07 | 0 | 0.02 | 0.04 |
| 0 | 0.03 | 0 | 0 | 0 | 0 | 0 | 0.01 |
| 0 | 0 | 0.04 | 0 | 0 | 0 | 0.01 | 0.01 |

Table S18: 2nd VB, EPC in eV/Å

| Γ | Z – *c* | Y - *b* | X - *a* | V | U | T | R |
|---|---|---|---|---|---|---|---|
| 0 | 0 | 0 | 0 | 0 | 0 | 0 | 0.27 |
| 0 | 0.1 | 0.26 | 0.02 | 0 | 0.06 | 0 | 0 |
| 0 | 0 | 0 | 0 | 0 | 0 | 0 | 0.12 |
| 0.07 | 0.02 | 0.05 | 0.06 | 0.2 | 0.01 | 0.19 | 0 |
| 0 | 0.06 | 0 | 0 | 0 | 0 | 0 | 0 |
| 0.03 | 0.02 | 0.28 | 0.01 | 0.04 | 0.06 | 0.11 | 0 |
| 0 | 0 | 0 | 0 | 0.23 | 0.11 | 0 | 0 |
| 0.01 | 0.01 | 0.15 | 0.05 | 0.05 | 0 | 0.07 | 0 |
| 0 | 0.01 | 0 | 0.09 | 0 | 0.14 | 0 | 0 |
| 0 | 0 | 0.08 | 0 | 0 | 0 | 0.06 | 0.06 |
| 0.01 | 0.02 | 0.05 | 0.01 | 0 | 0 | 0.01 | 0 |
| 0 | 0 | 0 | 0.35 | 0 | 0.04 | 0 | 0.12 |
| 0.01 | 0 | 0 | 0 | 0.06 | 0.09 | 0 | 0.03 |
| 0.31 | 0 | 0.23 | 0 | 0.26 | 0 | 0.04 | 0.06 |
| 0 | 0.05 | 0 | 0 | 0 | 0 | 0 | 0.15 |
| 0 | 0 | 0.09 | 0.02 | 0 | 0.25 | 0.2 | 0.19 |

Table S19: Davydov splitting, VB, EPC in eV/Å

| Γ | Z – *c* | Y - *b* | X - *a* | V | U | T | R |
|---|---|---|---|---|---|---|---|
| 0 | 0 | 0 | 0 | 0.01 | 0.01 | 0 | 0.45 |
| 0 | 0.17 | 0.48 | 0.08 | 0 | 0.17 | 0.2 | 0 |
| 0 | 0 | 0.01 | 0.01 | 0 | 0 | 0 | 0.47 |
| 0.07 | 0.03 | 0.07 | 0.02 | 0.29 | 0.21 | 0.47 | 0 |
| 0 | 0.14 | 0 | 0 | 0 | 0 | 0 | 0.05 |
| 0.02 | 0.12 | 0.33 | 0.02 | 0.18 | 0.06 | 0.15 | 0.28 |
| 0 | 0 | 0.01 | 0 | 0.07 | 0.15 | 0 | 0 |
| 0 | 0.02 | 0.26 | 0.18 | 0 | 0 | 0.01 | 0 |
| 0 | 0.05 | 0.01 | 0.19 | 0 | 0.24 | 0 | 0 |
| 0.01 | 0.01 | 0.01 | 0 | 0.01 | 0 | 0.15 | 0.01 |
| 0.02 | 0.01 | 0.01 | 0.18 | 0.01 | 0 | 0.04 | 0 |
| 0.01 | 0.04 | 0.01 | 0.06 | 0.01 | 0.01 | 0.01 | 0.24 |
| 0.05 | 0 | 0.01 | 0.01 | 0.16 | 0.03 | 0 | 0.16 |
| 0.01 | 0 | 0.02 | 0.01 | 0.31 | 0 | 0.12 | 0.11 |
| 0 | 0.16 | 0 | 0 | 0 | 0.01 | 0 | 0.07 |
| 0 | 0.01 | 0.11 | 0.02 | 0 | 0.05 | 0.11 | 0.03 |

Table S20: 1st CB, EPC in eV/Å

| Γ | Z – *c* | Y - *b* | X - *a* | V | U | T | R |
|---|---|---|---|---|---|---|---|
| 0 | 0 | 0 | 0 | 0 | 0 | 0 | 0.02 |
| 0 | 0.01 | 0.05 | 0.03 | 0 | 0.04 | 0.06 | 0 |
| 0 | 0 | 0 | 0 | 0 | 0 | 0 | 0.08 |
| 0.03 | 0.01 | 0.03 | 0.01 | 0 | 0.03 | 0.02 | 0 |
| 0 | 0.08 | 0 | 0 | 0 | 0 | 0 | 0 |
| 0 | 0.04 | 0.08 | 0.06 | 0.02 | 0 | 0.01 | 0.02 |
| 0 | 0 | 0 | 0 | 0.01 | 0.04 | 0 | 0 |
| 0.04 | 0.01 | 0.01 | 0.02 | 0.04 | 0 | 0 | 0 |
| 0 | 0.05 | 0 | 0.04 | 0 | 0.03 | 0 | 0 |
| 0 | 0 | 0.03 | 0 | 0 | 0 | 0.04 | 0.01 |
| 0.01 | 0.07 | 0.1 | 0.05 | 0 | 0 | 0.07 | 0 |
| 0 | 0 | 0 | 0.01 | 0 | 0.03 | 0 | 0.03 |
| 0.04 | 0 | 0 | 0 | 0.05 | 0.05 | 0 | 0.01 |
| 0.01 | 0 | 0.04 | 0 | 0.01 | 0 | 0.02 | 0.06 |
| 0 | 0.04 | 0 | 0 | 0 | 0 | 0 | 0.01 |
| 0 | 0 | 0.02 | 0.01 | 0 | 0.04 | 0 | 0.04 |

Table S21: 2nd CB, EPC in eV/Å

| Γ | Z – *c* | Y - *b* | X - *a* | V | U | T | R |
|---|---|---|---|---|---|---|---|
| 0 | 0.01 | 0.01 | 0 | 0.01 | 0.01 | 0 | 0.17 |
| 0 | 0.07 | 0.03 | 0.01 | 0 | 0.13 | 0.15 | 0 |
| 0 | 0 | 0.01 | 0.01 | 0 | 0 | 0 | 0.21 |
| 0.08 | 0.04 | 0.1 | 0.05 | 0.17 | 0.06 | 0.01 | 0.01 |
| 0 | 0.22 | 0.01 | 0.01 | 0.01 | 0 | 0 | 0.12 |
| 0.02 | 0.05 | 0.14 | 0 | 0.02 | 0.11 | 0.05 | 0.08 |
| 0 | 0.01 | 0.01 | 0 | 0.11 | 0.08 | 0.01 | 0 |
| 0 | 0.04 | 0.18 | 0.2 | 0.14 | 0.01 | 0.02 | 0 |
| 0 | 0.04 | 0 | 0 | 0.01 | 0.14 | 0 | 0 |
| 0 | 0.01 | 0.09 | 0.01 | 0.01 | 0.01 | 0.08 | 0.05 |
| 0.01 | 0.03 | 0.09 | 0.07 | 0.01 | 0 | 0.04 | 0 |
| 0.01 | 0.05 | 0.01 | 0.05 | 0.01 | 0.01 | 0.01 | 0.05 |
| 0.14 | 0 | 0.01 | 0.01 | 0.01 | 0.07 | 0.01 | 0.03 |
| 0.06 | 0 | 0.1 | 0.01 | 0.02 | 0.01 | 0.01 | 0.13 |
| 0 | 0.11 | 0.01 | 0 | 0 | 0 | 0 | 0.12 |
| 0 | 0 | 0.02 | 0.04 | 0 | 0.03 | 0.01 | 0.04 |

Table S22: Davydov splitting, CB, EPC in eV/Å

| Γ | Z – *c* | Y - *b* | X - *a* | V | U | T | R |
|---|---|---|---|---|---|---|---|
| 0.01 | 0.02 | 0.01 | 0.01 | 0.01 | 0.01 | 0.01 | 0.47 |
| 0.01 | 0.21 | 0.63 | 0.01 | 0.01 | 0.24 | 0.38 | 0.01 |
| 0.01 | 0.01 | 0.02 | 0.02 | 0 | 0.01 | 0.01 | 0.49 |
| 0.07 | 0 | 0.14 | 0.04 | 0.13 | 0.05 | 0.01 | 0.01 |
| 0.01 | 0.3 | 0.01 | 0.02 | 0.01 | 0.01 | 0.01 | 0.21 |
| 0.02 | 0.27 | 0.17 | 0.19 | 0.02 | 0.03 | 0.13 | 0.1 |
| 0.01 | 0.02 | 0.02 | 0.01 | 0.13 | 0.2 | 0.01 | 0.01 |
| 0.09 | 0.05 | 0.17 | 0.26 | 0.39 | 0.01 | 0.03 | 0.01 |
| 0.01 | 0.02 | 0.01 | 0.02 | 0.02 | 0.16 | 0.01 | 0.01 |
| 0.02 | 0.01 | 0.02 | 0.02 | 0.02 | 0.01 | 0.04 | 0.21 |
| 0.01 | 0.03 | 0.08 | 0.12 | 0.01 | 0.01 | 0.02 | 0.01 |
| 0.01 | 0.05 | 0.02 | 0.12 | 0.01 | 0.06 | 0.01 | 0 |
| 0.38 | 0.01 | 0.02 | 0.01 | 0.03 | 0.01 | 0.02 | 0.16 |
| 0.05 | 0.01 | 0.38 | 0.02 | 0.2 | 0.01 | 0.06 | 0.22 |
| 0.01 | 0.3 | 0.02 | 0.01 | 0.01 | 0.01 | 0.01 | 0.25 |
| 0.01 | 0.01 | 0.11 | 0.04 | 0 | 0.23 | 0.2 | 0.2 |

***Tetracene – film phase***

Intra-molecular modes, up to 400 cm$^{-1}$

Table S23: Frequencies in cm$^{-1}$

| Γ | Z – *c* | Y - *b* | X - *a* | V | U | T | R |
|---|---|---|---|---|---|---|---|
| 162.83 | 148.22 | 162.45 | 163.28 | 163.97 | 158.94 | 154.43 | 160.16 |
| 164.98 | 162.97 | 164.51 | 165.84 | 166.07 | 160.43 | 155.69 | 162.08 |
| 165.82 | 163.94 | 166.98 | 171.6 | 170.58 | 166.68 | 165.89 | 165.12 |
| 168.56 | 167.04 | 168.03 | 172.99 | 171.54 | 168.5 | 167.96 | 166.69 |
| 210.72 | 217.11 | 213.13 | 198.76 | 198.29 | 202.73 | 217.54 | 202.25 |
| 218.4 | 218.64 | 215.85 | 201.9 | 202.18 | 203.41 | 218.37 | 204.97 |
| 263.59 | 271.73 | 266.88 | 271.56 | 271.41 | 276.57 | 271.86 | 276.22 |
| 270.86 | 272.85 | 267.95 | 272.66 | 272.63 | 277.11 | 272.37 | 276.89 |
| 295.91 | 296.32 | 296.67 | 297.54 | 297.57 | 297.66 | 296.53 | 297.64 |
| 298.1 | 297.13 | 297.46 | 298.57 | 298.57 | 298.1 | 297 | 298.09 |
| 316.14 | 313.4 | 316.59 | 315.09 | 316.22 | 312.39 | 314.01 | 313.6 |
| 317.65 | 315.53 | 317.27 | 316.19 | 316.87 | 312.79 | 314.21 | 313.97 |
| 325.2 | 320.01 | 325.69 | 318.64 | 317.37 | 315.49 | 322.02 | 314.02 |
| 327.07 | 326.34 | 326.45 | 318.98 | 318.95 | 317.14 | 324.31 | 316.35 |
| 373.12 | 371.23 | 373.19 | 382.86 | 382.74 | 381.56 | 372.16 | 381.51 |
| 373.32 | 372.92 | 373.37 | 382.86 | 382.82 | 381.98 | 372.23 | 381.98 |

Table S24: 1$^{st}$ VB, EPC in eV/Å

| Γ | Z – *c* | Y - *b* | X - *a* | V | U | T | R |
|---|---|---|---|---|---|---|---|
| 0 | 0.01 | 0 | 0 | 0 | 0 | 0 | 0 |
| 0 | 0 | 0.01 | 0.04 | 0 | 0.03 | 0 | 0 |
| 0 | 0 | 0 | 0.01 | 0.02 | 0 | 0 | 0.02 |
| 0 | 0 | 0 | 0 | 0.04 | 0 | 0 | 0.01 |
| 0.05 | 0 | 0 | 0 | 0.01 | 0 | 0 | 0 |
| 0 | 0 | 0.05 | 0.04 | 0.04 | 0 | 0.04 | 0 |
| 0 | 0.01 | 0.01 | 0.02 | 0 | 0.02 | 0.02 | 0 |
| 0 | 0.01 | 0 | 0 | 0 | 0 | 0 | 0.03 |
| 0.01 | 0 | 0 | 0 | 0.03 | 0 | 0 | 0 |
| 0 | 0 | 0.01 | 0.01 | 0.01 | 0.06 | 0.04 | 0 |
| 0.01 | 0 | 0 | 0 | 0.06 | 0 | 0.06 | 0 |
| 0.01 | 0 | 0.01 | 0.01 | 0.02 | 0.03 | 0 | 0.02 |
| 0 | 0.05 | 0 | 0 | 0 | 0 | 0 | 0 |
| 0 | 0.01 | 0.04 | 0 | 0 | 0.03 | 0.05 | 0.04 |
| 0.01 | 0 | 0.01 | 0.01 | 0 | 0.01 | 0 | 0 |
| 0 | 0 | 0 | 0 | 0.02 | 0 | 0.03 | 0 |

Table S25: 2nd VB, EPC in eV/Å

| Γ | Z – *c* | Y - *b* | X - *a* | V | U | T | R |
|---|---|---|---|---|---|---|---|
| 0 | 0.23 | 0 | 0 | 0 | 0 | 0 | 0 |
| 0 | 0.03 | 0.09 | 0.07 | 0 | 0.12 | 0.1 | 0 |
| 0.02 | 0 | 0.02 | 0.03 | 0.08 | 0.05 | 0.04 | 0.07 |
| 0.01 | 0.07 | 0 | 0 | 0.06 | 0 | 0 | 0.06 |
| 0.07 | 0 | 0 | 0 | 0.08 | 0 | 0 | 0 |
| 0.11 | 0 | 0.18 | 0.17 | 0.14 | 0 | 0.07 | 0 |
| 0 | 0.01 | 0 | 0.03 | 0 | 0 | 0 | 0.03 |
| 0 | 0 | 0 | 0 | 0 | 0 | 0 | 0.05 |
| 0.03 | 0 | 0 | 0 | 0.04 | 0 | 0 | 0 |
| 0 | 0 | 0.02 | 0.03 | 0.03 | 0.04 | 0.02 | 0 |
| 0 | 0 | 0 | 0 | 0.31 | 0 | 0.14 | 0 |
| 0.01 | 0 | 0.03 | 0.01 | 0.06 | 0.13 | 0 | 0.16 |
| 0 | 0.06 | 0 | 0 | 0 | 0 | 0 | 0 |
| 0 | 0.24 | 0.03 | 0.01 | 0 | 0.23 | 0.33 | 0.29 |
| 0 | 0 | 0 | 0 | 0.02 | 0.12 | 0 | 0 |
| 0.02 | 0 | 0 | 0 | 0.04 | 0 | 0.08 | 0 |

Table S26: Davydov splitting, VB, EPC in eV/Å

| Γ | Z – *c* | Y - *b* | X - *a* | V | U | T | R |
|---|---|---|---|---|---|---|---|
| 0 | 0.02 | 0.01 | 0 | 0.01 | 0.01 | 0 | 0 |
| 0 | 0.04 | 0.22 | 0.12 | 0.01 | 0.17 | 0.01 | 0.01 |
| 0.01 | 0 | 0.09 | 0.02 | 0.14 | 0.09 | 0.06 | 0.41 |
| 0.02 | 0.04 | 0.01 | 0.01 | 0.17 | 0 | 0 | 0.16 |
| 0.27 | 0 | 0 | 0 | 0.07 | 0 | 0 | 0 |
| 0.06 | 0.01 | 0.28 | 0.24 | 0.22 | 0 | 0.14 | 0.01 |
| 0 | 0.02 | 0.05 | 0.05 | 0 | 0.02 | 0.01 | 0.04 |
| 0 | 0 | 0 | 0 | 0 | 0 | 0.01 | 0.1 |
| 0.06 | 0.01 | 0.01 | 0.01 | 0.13 | 0.01 | 0 | 0 |
| 0 | 0 | 0.05 | 0.05 | 0.06 | 0.01 | 0.07 | 0 |
| 0.01 | 0 | 0 | 0 | 0.49 | 0 | 0.85 | 0 |
| 0.05 | 0 | 0.06 | 0.04 | 0.14 | 0.07 | 0 | 0.1 |
| 0 | 0.27 | 0 | 0 | 0 | 0.01 | 0.01 | 0 |
| 0 | 0.09 | 0.14 | 0.01 | 0 | 0.23 | 0.35 | 0.27 |
| 0.02 | 0 | 0.02 | 0.04 | 0.01 | 0.03 | 0 | 0.01 |
| 0.01 | 0.01 | 0 | 0 | 0.07 | 0 | 0.11 | 0.01 |

Table S27: 1st CB, EPC in eV/Å

| Γ | Z – *c* | Y - *b* | X - *a* | V | U | T | R |
|---|---|---|---|---|---|---|---|
| 0 | 0.01 | 0 | 0 | 0 | 0 | 0 | 0 |
| 0 | 0.06 | 0.04 | 0.03 | 0 | 0.02 | 0.13 | 0 |
| 0.02 | 0 | 0.09 | 0.11 | 0.07 | 0.04 | 0.07 | 0.01 |
| 0.13 | 0.01 | 0 | 0 | 0.05 | 0 | 0 | 0.05 |
| 0.01 | 0 | 0 | 0 | 0.07 | 0 | 0 | 0 |
| 0.01 | 0 | 0.02 | 0.03 | 0.02 | 0.05 | 0 | 0 |
| 0 | 0.03 | 0.07 | 0.04 | 0 | 0.08 | 0.09 | 0.06 |
| 0 | 0.08 | 0 | 0 | 0 | 0 | 0 | 0.06 |
| 0.01 | 0 | 0 | 0 | 0.01 | 0 | 0 | 0 |
| 0 | 0 | 0 | 0.01 | 0 | 0.08 | 0.04 | 0 |
| 0.03 | 0 | 0 | 0 | 0.02 | 0 | 0.03 | 0 |
| 0.03 | 0 | 0.01 | 0.03 | 0 | 0.04 | 0 | 0.11 |
| 0 | 0.01 | 0 | 0 | 0 | 0 | 0 | 0 |
| 0 | 0.03 | 0.02 | 0.06 | 0 | 0 | 0.03 | 0.03 |
| 0.02 | 0 | 0.02 | 0.03 | 0.02 | 0.03 | 0 | 0 |
| 0.02 | 0 | 0 | 0 | 0.03 | 0 | 0.01 | 0 |

Table S28: 2nd CB, EPC in eV/Å

| Γ | Z – *c* | Y - *b* | X - *a* | V | U | T | R |
|---|---|---|---|---|---|---|---|
| 0 | 0.07 | 0.01 | 0 | 0.01 | 0.01 | 0.01 | 0 |
| 0 | 0.17 | 0.04 | 0.13 | 0.01 | 0.11 | 0.22 | 0.01 |
| 0.02 | 0 | 0.07 | 0.1 | 0.03 | 0.15 | 0.21 | 0.08 |
| 0.11 | 0.04 | 0.01 | 0.01 | 0.08 | 0 | 0 | 0.17 |
| 0.01 | 0 | 0 | 0.01 | 0.12 | 0 | 0 | 0 |
| 0.05 | 0.01 | 0.05 | 0.06 | 0.03 | 0.01 | 0.03 | 0.01 |
| 0.01 | 0.01 | 0.04 | 0.09 | 0 | 0.03 | 0.06 | 0.01 |
| 0 | 0.06 | 0.01 | 0 | 0.01 | 0 | 0 | 0.07 |
| 0.02 | 0 | 0.01 | 0.01 | 0 | 0.01 | 0 | 0 |
| 0.01 | 0 | 0.02 | 0.03 | 0.01 | 0.08 | 0.06 | 0 |
| 0.16 | 0 | 0 | 0.01 | 0.14 | 0 | 0.24 | 0.01 |
| 0.02 | 0.01 | 0.07 | 0.07 | 0.11 | 0.04 | 0 | 0.25 |
| 0 | 0.05 | 0.01 | 0 | 0 | 0.01 | 0 | 0 |
| 0 | 0.09 | 0.07 | 0.01 | 0.01 | 0.01 | 0.05 | 0.03 |
| 0 | 0 | 0 | 0.02 | 0 | 0.14 | 0 | 0.01 |
| 0.02 | 0 | 0.01 | 0 | 0.03 | 0.01 | 0.09 | 0 |

Table S29: Davydov splitting, CB, EPC in eV/Å

| Γ | Z – *c* | Y - *b* | X - *a* | V | U | T | R |
|---|---|---|---|---|---|---|---|
| 0.01 | 0.01 | 0.02 | 0.01 | 0.01 | 0.02 | 0.01 | 0.01 |
| 0 | 0.35 | 0.11 | 0.12 | 0.01 | 0.18 | 0.32 | 0.02 |
| 0.02 | 0 | 0 | 0.02 | 0.02 | 0.22 | 0.35 | 0.28 |
| 0.06 | 0.04 | 0.01 | 0.02 | 0.04 | 0.01 | 0 | 0.34 |
| 0.05 | 0.01 | 0.01 | 0.01 | 0.11 | 0.01 | 0.01 | 0.01 |
| 0.01 | 0.01 | 0.03 | 0.02 | 0.05 | 0.02 | 0 | 0.01 |
| 0.01 | 0.06 | 0.09 | 0.02 | 0.01 | 0.11 | 0.04 | 0.06 |
| 0.01 | 0 | 0.02 | 0 | 0.01 | 0 | 0.01 | 0.03 |
| 0.05 | 0.01 | 0.01 | 0.02 | 0.38 | 0.02 | 0 | 0.01 |
| 0.02 | 0.01 | 0.03 | 0.06 | 0.02 | 0.02 | 0.24 | 0.01 |
| 0.11 | 0.01 | 0.01 | 0.02 | 0.66 | 0.01 | 0.45 | 0.02 |
| 0.15 | 0.02 | 0.16 | 0.07 | 0.07 | 0.06 | 0 | 0.37 |
| 0.01 | 0.21 | 0.01 | 0 | 0 | 0.02 | 0.01 | 0.01 |
| 0 | 0.01 | 0.1 | 0.11 | 0.01 | 0.18 | 0.21 | 0.21 |
| 0.02 | 0.01 | 0.02 | 0.02 | 0.01 | 0.23 | 0.01 | 0.01 |
| 0.04 | 0.01 | 0.01 | 0.01 | 0.09 | 0.02 | 0.19 | 0.01 |

Table S30: DOS-weighted average EPC as used for the charge transport calculation, for the P1 tetracene polymorphs. The EPC are in units of eV/Å.

| $\hbar\omega$ [meV] | intra- VB1 | intra- VB2 | inter- VB | intra- CB1 | intra- CB2 | inter- CB |
|---|---|---|---|---|---|---|
| 3.50 | 0.08 | 0.08 | 0.13 | 0.13 | 0.24 | 0.18 |
| 4.01 | 0.25 | 0.25 | 0.10 | 0.25 | 0.48 | 0.65 |
| 5.09 | 0.18 | 0.23 | 0.03 | 0.23 | 0.35 | 0.60 |
| 5.46 | 0.07 | 0.09 | 0.11 | 0.06 | 0.11 | 0.20 |
| 5.97 | 0.08 | 0.14 | 0.10 | 0.12 | 0.16 | 0.23 |
| 6.63 | 0.10 | 0.06 | 0.07 | 0.08 | 0.10 | 0.14 |
| 7.78 | 0.07 | 0.09 | 0.06 | 0.08 | 0.08 | 0.13 |
| 8.41 | 0.10 | 0.14 | 0.16 | 0.09 | 0.19 | 0.22 |
| 9.89 | 0.05 | 0.04 | 0.09 | 0.04 | 0.09 | 0.07 |
| 10.39 | 0.07 | 0.07 | 0.22 | 0.06 | 0.10 | 0.16 |
| 11.50 | 0.04 | 0.07 | 0.26 | 0.07 | 0.10 | 0.09 |
| 11.97 | 0.03 | 0.03 | 0.13 | 0.04 | 0.07 | 0.07 |
| 13.13 | 0.04 | 0.10 | 0.19 | 0.08 | 0.08 | 0.21 |
| 14.52 | 0.04 | 0.12 | 0.48 | 0.07 | 0.07 | 0.19 |
| 16.10 | 0.04 | 0.07 | 0.21 | 0.06 | 0.09 | 0.10 |
| 16.58 | 0.04 | 0.12 | 0.45 | 0.08 | 0.05 | 0.16 |

Table S31: DOS-weighted average EPC as used for the charge transport calculation, for the PF tetracene polymorphs. The EPC are in units of eV/Å.

| $\hbar\omega$ [meV] | intra- VB1 | intra- VB2 | inter- VB | intra- CB1 | intra- CB2 | inter- CB |
|---|---|---|---|---|---|---|
| 3.55 | 0.03 | 0.13 | 0.21 | 0.01 | 0.08 | 0.22 |
| 3.75 | 0.03 | 0.10 | 0.22 | 0.04 | 0.09 | 0.31 |
| 4.80 | 0.02 | 0.04 | 0.16 | 0.03 | 0.07 | 0.17 |
| 5.06 | 0.03 | 0.11 | 0.22 | 0.02 | 0.08 | 0.08 |
| 5.61 | 0.01 | 0.02 | 0.06 | 0.03 | 0.10 | 0.14 |
| 6.15 | 0.05 | 0.12 | 0.19 | 0.04 | 0.08 | 0.14 |
| 8.23 | 0.02 | 0.09 | 0.06 | 0.01 | 0.05 | 0.08 |
| 8.90 | 0.01 | 0.06 | 0.11 | 0.02 | 0.11 | 0.20 |
| 10.20 | 0.03 | 0.06 | 0.11 | 0.03 | 0.05 | 0.05 |
| 10.78 | 0.02 | 0.04 | 0.04 | 0.01 | 0.04 | 0.08 |
| 11.61 | 0.01 | 0.02 | 0.07 | 0.05 | 0.04 | 0.06 |
| 12.05 | 0.02 | 0.12 | 0.10 | 0.02 | 0.03 | 0.05 |
| 13.63 | 0.02 | 0.04 | 0.09 | 0.03 | 0.04 | 0.12 |
| 14.16 | 0.03 | 0.19 | 0.14 | 0.02 | 0.06 | 0.17 |
| 16.11 | 0.01 | 0.05 | 0.06 | 0.01 | 0.05 | 0.12 |
| 16.02 | 0.01 | 0.17 | 0.05 | 0.03 | 0.03 | 0.16 |

**C – convergence in the EPC**

The convergence tests of the EPC in respect of the atoms displacement, eq.s (1) and (2) of the main text, focused on the topmost VB of the P1 polymorph of tetracene. These tests are reported in Figure S1. Theoretically the displacement should tend to zero, however we observed a nearly linear increase of the EPC with the decrease of the displacement amount; we chose to use the displacement at which this increase stopped. As an overall estimation of the EPC, we used the sum of all the EPC, denoted as EPC* in Figure S1.

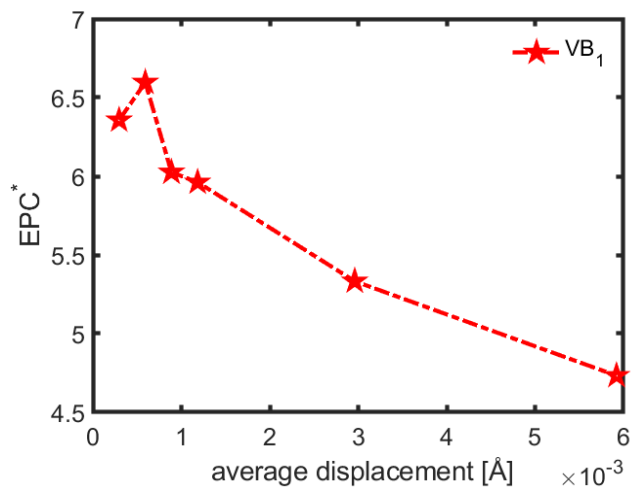

Figure S1: convergence test on the EPC* (sum of the individual EPC values) in respect of the average atoms displacement, performed on the topmost VB ($VB_1$) of the tetracene bulk polymorph 1. At around 0.001 Å the EPC* appeared to saturate. For smaller displacements, numerical noises started to appear.

## D – Deformation potentials for naphthalene, anthracene, and ADP

For the sake of completeness, we report here the DOS-weighted average of the mode and *q*-point dependent EPC values and related frequency, for naphthalene, Table S32, and anthracene, Table S33. The Table S34 reports the ADP values obtained with eq.s (5) and (6) on the oligoacenes investigated. For comparison, we report also the ADP values reported in literature as obtained with the method of the relative shift of the band edge.[1]

Table S32: DOS-weighted average of the mode and *q*-point dependent phonon frequency and EPC values for naphthalene.

| $\hbar\omega$ [meV] | intra- VB1 | intra- VB2 | inter- VB | intra- CB1 | intra- CB2 | inter- CB |
|---|---|---|---|---|---|---|
| 5.17 | 0.25 | 0.20 | 0.10 | 0.07 | 0.18 | 0.39 |
| 5.18 | 0.03 | 0.03 | 0.04 | 0.01 | 0.11 | 0.02 |
| 6.44 | 0.20 | 0.28 | 0.35 | 0.05 | 0.08 | 0.18 |
| 6.47 | 0.18 | 0.17 | 0.18 | 0.05 | 0.14 | 0.22 |
| 7.24 | 0.06 | 0.06 | 0.09 | 0.04 | 0.07 | 0.12 |
| 7.38 | 0.10 | 0.09 | 0.11 | 0.06 | 0.16 | 0.24 |
| 8.90 | 0.35 | 0.40 | 0.47 | 0.08 | 0.15 | 0.39 |
| 9.06 | 0.15 | 0.19 | 0.24 | 0.03 | 0.17 | 0.30 |
| 11.14 | 0.04 | 0.10 | 0.23 | 0.06 | 0.18 | 0.45 |
| 11.46 | 0.09 | 0.08 | 0.18 | 0.05 | 0.15 | 0.27 |
| 13.97 | 0.04 | 0.04 | 0.08 | 0.04 | 0.03 | 0.14 |
| 13.92 | 0.09 | 0.09 | 0.10 | 0.02 | 0.05 | 0.11 |

Table S33: DOS-weighted average of the mode and *q*-point dependent phonon frequency and EPC values for anthracene.

| $\hbar\omega$ [meV] | intra- VB1 | intra- VB2 | inter- VB | intra- CB1 | intra- CB2 | inter- CB |
|---|---|---|---|---|---|---|
| 4.35 | 0.11 | 0.16 | 0.28 | 0.16 | 0.24 | 0.42 |
| 4.36 | 0.07 | 0.10 | 0.18 | 0.17 | 0.22 | 0.40 |
| 5.60 | 0.10 | 0.17 | 0.33 | 0.18 | 0.15 | 0.31 |
| 5.55 | 0.06 | 0.18 | 0.37 | 0.15 | 0.18 | 0.36 |
| 6.23 | 0.13 | 0.23 | 0.44 | 0.22 | 0.22 | 0.44 |
| 6.35 | 0.06 | 0.14 | 0.24 | 0.15 | 0.17 | 0.31 |
| 7.88 | 0.06 | 0.16 | 0.31 | 0.10 | 0.13 | 0.24 |
| 8.24 | 0.03 | 0.05 | 0.14 | 0.07 | 0.10 | 0.20 |
| 10.01 | 0.05 | 0.13 | 0.29 | 0.10 | 0.13 | 0.28 |
| 10.28 | 0.06 | 0.15 | 0.32 | 0.15 | 0.14 | 0.30 |
| 13.27 | 0.08 | 0.11 | 0.22 | 0.08 | 0.08 | 0.12 |
| 13.59 | 0.06 | 0.11 | 0.22 | 0.09 | 0.10 | 0.24 |
| 15.62 | 0.09 | 0.12 | 0.21 | 0.10 | 0.09 | 0.22 |
| 15.79 | 0.12 | 0.15 | 0.23 | 0.10 | 0.08 | 0.18 |

Table S34: ADP deformation potential values for naphthalene, anthracene, and tetracene P1, computed with the method in eq.s (5) and (6), and comparison with literature values computed with a different method, which have been computed for unit cell dilation/compression along the *a* and *b* axes, and the corresponding values are indicating with a superscript *a* or *b*, accordingly.[1] Values in eV. We observe that are generally very small numbers if compared with inorganic semiconductors,[2] except for the values of one of the valence band and for the inter-band in the CB in anthracene, which are unphysically high, a point that deserves further investigation.

|  | naphthalene | | anthracene | | tetracene P1 | |
|---|---|---|---|---|---|---|
| $VB_1$ | 0.24 | $1.31^a$ | 0.70 | $1.12^a$ | 0.17 | $1.79^a$ |
| $VB_2$ | 0.38 | $1.39^b$ | 6.32 | $1.38^b$ | 0.16 | $0.47^b$ |
| $VB_{inter-band}$ | 0.32 | | 7.39 | | 0.76 | |
| $CB_1$ | 0.17 | $0.96^a$ | 0.95 | $0.42^a$ | 0.25 | $1.6^a$ |
| $CB_2$ | 0.00 | $0.56^b$ | 0.00 | $0.87^b$ | 0.00 | $0.53^b$ |
| $CB_{inter-band}$ | 0.84 | | 20.39 | | 0.69 | |

# E – Relaxation times

In this section we report the relaxation times along the *a*, *b*, and *c* directions for the hole transport in the valence band of naphthalene, anthracene, and tetracene, along with the corresponding density of states (DOS). These are reported in Figure S2 for the intra- (blue) and inter- (green) band processes, and the DOS related to the two VB bands are reported with a different tones of red color. The zero of the energy corresponds to the VB edge and the positive energy are referred to the hole transport state energy, i.e. they are into the VB. For the case of naphthalene, we report also the case of ADP scattering with acoustic phonon at zone center (red). To highlight at what extent it is larger, and hence negligible. The insets in the naphthalene cases report the same curves without the ADP.

We can note that for naphthalene the intra- and inter- band processes have similar strength, while for anthracene the inter-band processes are much stronger, resulting in shorter relaxation time. In tetracene, the inter-band scattering is negligible, with a relaxation time not defined, set to zero for graphical purposes. This is because in tetracene the two bands are quite split apart with nearly absent overlap, as observable in the DOS in Figure S2 (n), while in naphthalene and anthracene the two bands overlap for the largest part of the energy of interest, thus the inter-band scattering processes are allowed throughout the entire hole energy range of interest in transport. We observe that the relaxation times are larger at the band edge, where only phonon absorption is allowed and the final available DOS is small, and suddenly decreases as soon as the DOS increases and phonon emission is possible.

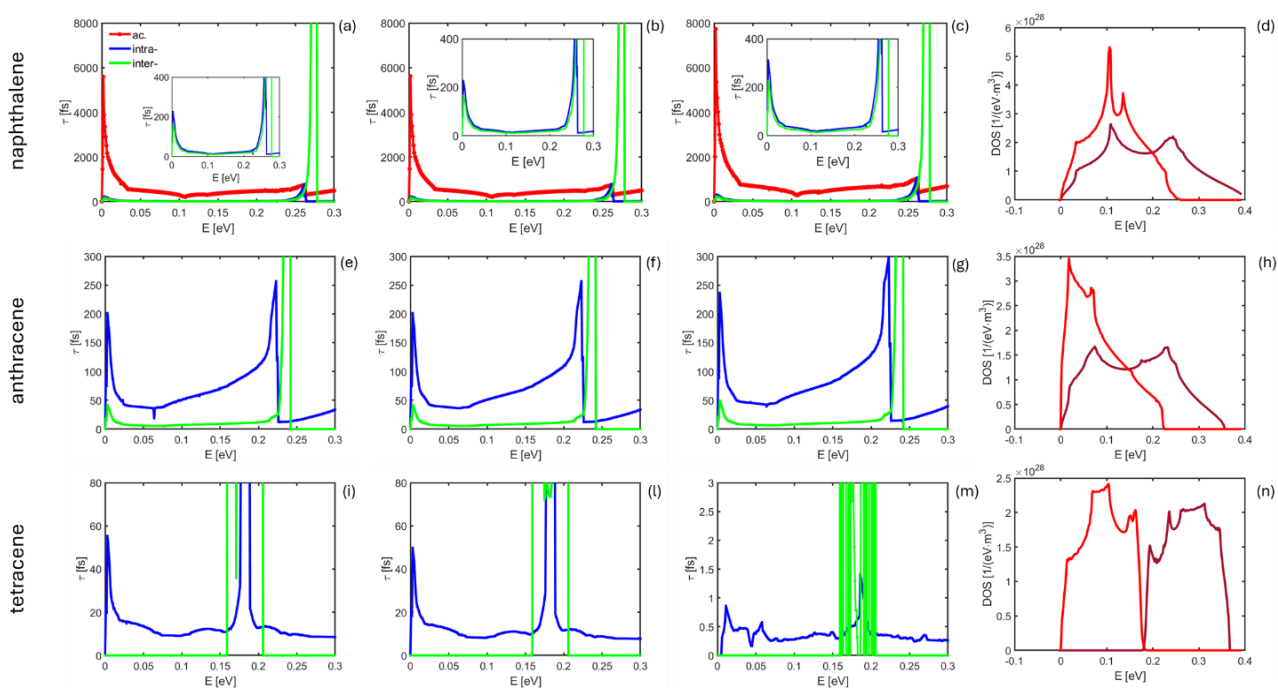

Figure S2: relaxation times for naphthalene (a-c), anthracene (e-g), tetracene(i-m) for the transport along the *a* (a, e, i), *b* (b, f, l), and *c* (c, g, m) axes. In blue and green are depicted the relaxation times for intra- and inter- band

processes, respectively, while the red lines in (a-c) represent the relaxation times for ADP mechanism. The corresponding density-of-states (DOS) are reported in (d), (h) and (n), the two VB are depicted with two red tones. The zero of the energy corresponds to the VB edge and the positive energy are referred to the hole transport state energy, i.e. they are into the VB.